# High-yield large-scale suspended graphene membranes over closed cavities for sensor applications

*Sebastian Lukas[1], Ardeshir Esteki[1], Nico Rademacher[2], Vikas Jangra[1], Michael Gross[1], Zhenxing Wang[2], Ha-Duong Ngo[3], Manuel Bäuscher[4], Piotr Mackowiak[4], Katrin Höppner[4], Dominique J. Wehenkel[5], Richard van Rijn[5], Max C. Lemme*[1,2]*

[1]Chair of Electronic Devices, RWTH Aachen University, Otto-Blumenthal-Str. 25, 52074 Aachen, Germany

[2]AMO GmbH, Advanced Microelectronic Center Aachen, Otto-Blumenthal-Str. 25, 52074 Aachen, Germany

[3]University of Applied Sciences Berlin, Wilhelminenhofstr. 75A (C 525), 12459 Berlin, Germany

[4]Fraunhofer IZM, Gustav-Meyer-Allee 25, 13355 Berlin, Germany

[5]Applied Nanolayers B.V., Feldmannweg 17, 2628 CT Delft, The Netherlands

* max.lemme@eld.rwth-aachen.de

## Abstract

Suspended membranes of monatomic graphene exhibit great potential for applications in electronic and nanoelectromechanical devices. In this work, a "hot and dry" transfer process is demonstrated to address the fabrication and patterning challenges of large-area graphene membranes on top of closed, sealed cavities. Here, "hot" refers to the use of high temperature during transfer, promoting the adhesion. Additionally, "dry" refers to the absence of liquids when graphene and target substrate are brought into contact. The method leads to higher yields of intact suspended monolayer CVD graphene and artificially stacked double-layer CVD graphene membranes than previously reported. The yield evaluation is performed using neural-network-based object detection in SEM images, ascertaining high yields of intact membranes with large statistical accuracy. The suspended membranes are examined by Raman tomography and AFM. The method is verified by applying the suspended graphene devices as piezoresistive pressure sensors. Our technology advances the application of suspended graphene membranes and can be extended to other two-dimensional materials.

Graphene, a two-dimensional (2D) material consisting of a single atomic layer of carbon atoms, has received significant attention in recent years. Apart from its intriguing electronic properties, [1] graphene's mechanical strength [2,3] and extraordinary hermeticity [4–6] have made it a promising candidate for micro- and nanoelectromechanical systems (MEMS and NEMS) involving freely suspended membranes of graphene. [7,8] Very strong adhesion of monolayer graphene to $SiO_2$ substrates [9] enables robust monolayer graphene membranes for long-lasting NEMS. Suspending graphene allows the construction of highly sensitive pressure sensors, [10–15] microphones, [16–20] accelerometers, [21–23] and mass and gas sensors. [24–28]

When targeting industrial applications of graphene-based sensors, scalable fabrication methods must be used. Many high-performance graphene devices have been demonstrated based on exfoliated single-crystal graphene of micrometer dimensions. However, scalable approaches like wafer-size chemical vapor deposition (CVD) [29,30] or graphene dispersions [31] for spin coating and inkjet printing [32,33] will be required for future industrial fabrication.

This work focuses on sensor applications that require closed cavities sealed by a thin graphene membrane, such as NEMS pressure sensors for measuring absolute pressure. For such applications, under-etching of directly-grown or transferred graphene by means of hydrogen fluoride (HF) vapor, as demonstrated for open-cavity suspended graphene, [17,34] is not feasible. Instead, different wet [35,36] and dry [35,37] transfer methods have been utilized previously, with varying levels of success. Commonly used wet transfer methods often lead to membrane ruptures due to the liquid inside the cavity, which pulls the membrane down and collapses it during drying. Dry transfer methods avoid this problem, but those involving thermal release tape or polydimethylsiloxane (PDMS) stamps have also proven unsuitable due to high mechanical stress during the transfer, inducing cracks in the graphene. Several works have used supporting polymer layers on top of the graphene membranes to increase mechanical stability. [34,38–41] Such concepts benefit some applications but generally result in lower sensitivity or larger device footprints. [42,43]

Here, we have developed a frame-based temperature-assisted dry transfer method to for monolayer CVD graphene and artificially stacked double-layer CVD graphene onto target substrates with pre-etched cavities. [44] We applied this transfer method to both in-house-grown and commercially available CVD graphene. Additionally, we investigated the removal of polymethylmethacrylate (PMMA) from the graphene membranes using liquid solvents or annealing in an inert atmosphere. Furthermore, we compared our transfer method to a wafer-scale dry transfer of double-layer CVD graphene using a proprietary commercially available transfer method. Atomic force microscopy (AFM) and Raman spectroscopy confirm that our membranes are freely suspended. The yield of intact membranes after transfer, patterning, and PMMA removal was evaluated by automated scanning electron microscopy (SEM) image acquisition and processing using TensorFlow object detection, resulting in statistics with large quantities of membranes (over 2,000,000 across all evaluated samples) for varying cavity diameters. The high number of examined membranes makes this work highly statistically relevant. Finally, graphene-membrane-based devices were measured in a pressure chamber to demonstrate their performance as piezoresistive pressure sensors.

## Results

Target substrates with cavities etched into the silicon (Si) substrates, then passivated with thermally grown silicon dioxide ($SiO_2$), and equipped with bottom metal contacts, were fabricated as described in detail in the Methods section. Single-layer and double-layer graphene was then transferred onto 2 × 2 cm² chips over the closed cavities to form suspended membranes, and patterned using a specific photoresist stack. The process is schematically presented in **Figure 1**a-j. Photographs of the graphene frames before and during transfer are shown in **Figure 1**k-n. Furthermore, graphene was transferred in the same way onto wafer-scale cavity substrates prepared by Fraunhofer IZM, featuring arrays of cavities with larger spacing.

**Figure 1**m shows a 150 mm wafer from the same fabrication flow as the 2 × 2 cm² chips, with wafer-scale-transferred graphene using a proprietary method of Applied Nanolayers B.V., Netherlands (ANL). This wafer was additionally used for graphene membrane fabrication and yield analysis. The results retrieved from this wafer are included here to report on the commercial availability and suitability of wafer-scale graphene transfer for membrane-based applications.

An overview of all fabricated and evaluated samples is provided in **Table 1**. While the patterning process with the three-layer resist stack was the same for all samples, there were differences in the graphene source, layer number, transfer process, and PMMA removal process.

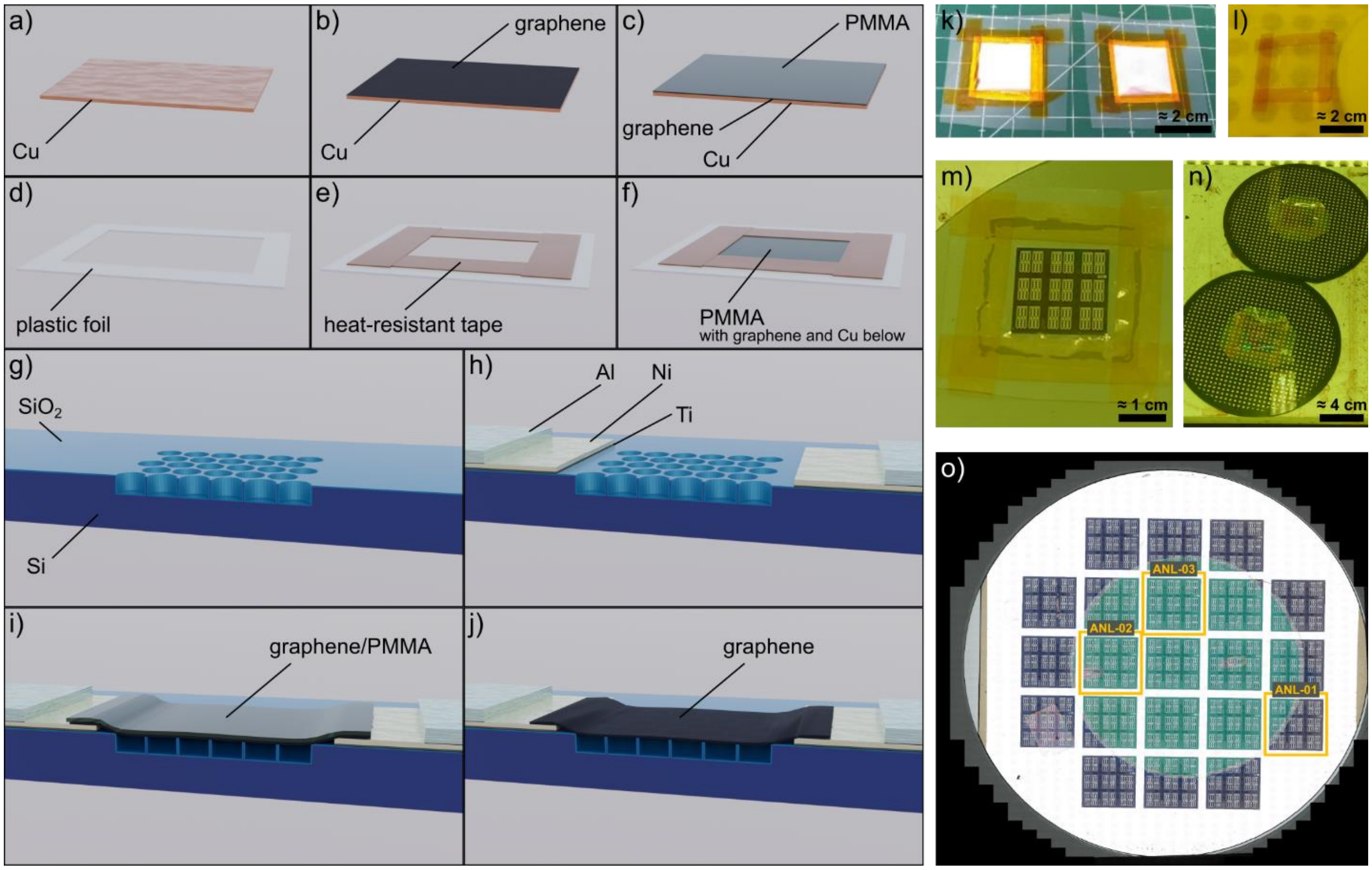

**Figure 1.** Schematics and photos of the fabrication process and device structure. (a) Cu foil before graphene growth. (b) Cu foil with grown graphene. (c) Cu foil with graphene and spin-coated PMMA. (d) Plastic frame for transfer. (e) Plastic frame with heat-resistant tape. (f) Transfer frame with tape attached to PMMA/graphene/Cu foil stack. (g) Cross-section of Si/$SiO_2$ substrate with etched cavities and (h) after Ni and Al contact deposition and patterning. (i) Cross-section of device after transfer and patterning of graphene/PMMA stack and (j) after PMMA removal with suspended graphene across cavities. (k) Photograph of transfer frames with PMMA/graphene/Cu foil attached. (l) Photograph of transfer frame after Cu etching during drying. (m-n) Photographs of transfer frames with PMMA/graphene and target substrates (2 × 2 $cm^2$ chip and 150 mm wafers, respectively) during hot and dry transfer process. (o) Stitched micrograph of 150 mm wafer after 4’’ graphene transfer by ANL.

**Table 1.** Overview of the fabricated and analyzed samples. The naming convention in this work is [graphene source]-[layers]-[PMMA removal process]-[sample number], *e.g.*, "inH-G2-A-2321" for a sample with double-layer in-house-grown graphene and PMMA removed by annealing.

| graphene source | graphene type | transfer process | PMMA removal process | sample name |
|---|---|---|---|---|
| commercial (CVD graphene on Cu foil) | artificially stacked double layer | hot & dry, frame-based, up to 3 cm × 3 cm graphene | annealing | com-G2-A-IZM5<br>com-G2-A-IZM6 |
| | | | annealing | com-G2-A-2b<br>com-G2-A-2d<br>com-G2-A-2311<br>com-G2-A-2312 |
| | | | liquid solvent | com-G2-S-2a<br>com-G2-S-2c |
| | single layer | hot & dry, frame-based, up to 3 cm × 3 cm graphene | annealing | com-G1-A-1b<br>com-G1-A-2b<br>com-G1-A-2316<br>com-G1-A-2317<br>com-G1-A-2318<br>com-G1-A-2323<br>com-G1-A-2324 |
| | | | liquid solvent | com-G1-S-1a<br>com-G1-S-2a<br>com-G1-S-2c |
| in-house-grown (CVD graphene on Cu foil) | artificially stacked double layer | hot & dry, frame-based, up to 3 cm × 3 cm graphene | annealing | inH-G2-A-2321<br>inH-G2-A-2322 |
| | single layer | | annealing | inH-G1-A-2331<br>inH-G1-A-2332 |
| Applied Nanolayers, B.V. | artificially stacked double layer | commercial transfer service, proprietary (dry, wafer-bonding-based), 4'' graphene | annealing | ANL-G2-A-1<br>ANL-G2-A-2<br>ANL-G2-A-3 |

The frame-based hot and dry transfer process generally resulted in a high yield of intact single and double-layer CVD graphene membranes across closed cavities of a few µm depth. **Figure 2**a-b shows two SEM image examples of the graphene membranes before software processing and cut-outs show parts of the images after automated detection and labelling of broken (red) and intact (green) membranes. More SEM images are shown in **Figure S1**. The detection algorithm detected the majority of intact and broken membranes. We estimate the amount of non-detected or wrongly classified membranes (broken instead of intact, or vice versa) to be below 1 % for the small membranes up to 3.4 µm diameter and below 5 % for the larger membranes, based on randomly chosen images and manual counting. The correctness of this estimate can unfortunately not be guaranteed. Since the automated detection allowed the classification of more than 2,000,000 membranes across all samples, the significantly greater sample size compared to manual counting and classifying compensates for the small uncertainty in detection. The number of detected membranes varies by sample group due to the varying number of samples per sample group and the varying size of the evaluated sample area, which is based on the size of the transferred graphene pieces (**Figure 2**c). The number of detected membranes also varies by membrane diameter, which reflects the distribution of cavities and their diameter included in the layout. All samples contained membranes with six different diameters (1.5 µm, 2.3 µm, 3.4 µm, 5.2 µm, 7.9 µm, and 12 µm), except for the samples on Fraunhofer IZM substrates, which only contained membranes with three different diameters (1.5 µm, 2.0 µm, and 3 µm).

We observed a difference after the resist removal step (after graphene channel patterning) between the in-house-transferred graphene and the commercial transfer service by ANL. The commercially transferred graphene delaminated in several locations of the samples, leaving some cavity arrays entirely uncovered, while adhesion was not an issue for the in-house-transferred graphene (**Figure S2**). Arrays of larger cavities were more affected by this problem than those of smaller cavities. Entirely uncovered cavity arrays on the commercially transferred graphene samples were excluded from the membrane yield analysis.

**Figure 2**d shows the ratio of intact versus broken membranes for the various samples as a function of the cavity diameter. The yield of intact membranes generally decreases with increasing

membrane diameter. The best samples, *i.e.*, the top values of the displayed error bars, reach up to 99 % yield of intact graphene membranes. The maximum yields for the respective graphene types and processing variations are summarized in **Table S1**.

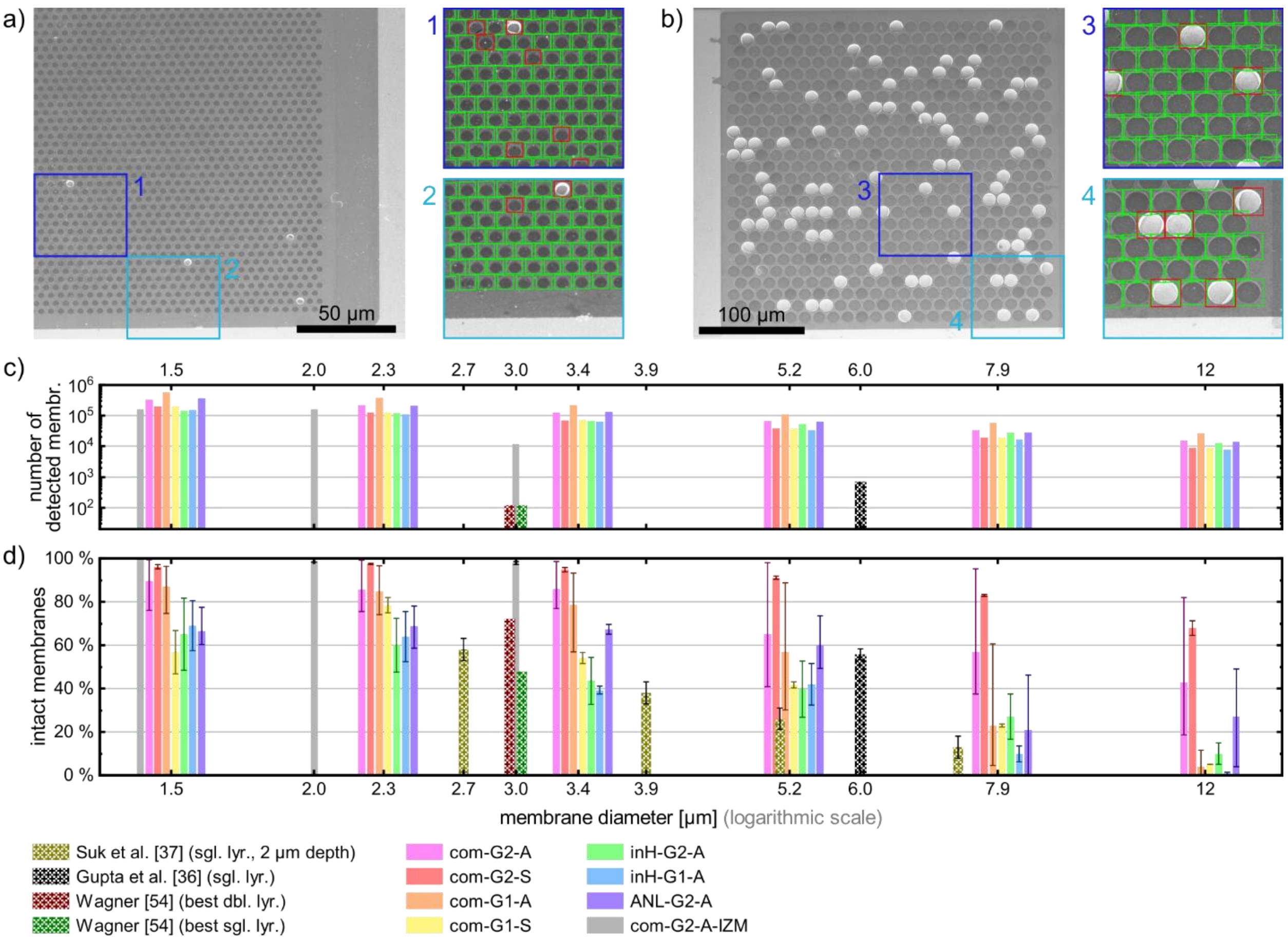

**Figure 2.** (a, b) Two example SEM images of graphene membrane arrays with cut-outs after automated image processing and yield analysis. The object detection algorithm marks intact graphene membranes with green frames and (partially) broken membranes with red frames. (c) Total number of detected membranes by sample group. (d) Ratio of intact membranes and number of detected membranes by sample group. The error bars show the maximum and minimum yields for the different analyzed samples, while the bar height shows the mean value. Data (incl. error bar values) is provided in **Table S1**.

The evaluation shows that double-layer graphene generally leads to a higher yield of intact membranes than single-layer graphene, in line with previous reports. [22,35] This is particularly evident in large diameter membranes (7.9 µm and 12 µm), where the yield for single-layer graphene membranes drops significantly. The increased mechanical stability of double-layer graphene over that of single-layer graphene is primarily influenced by the grain boundaries in the graphene monolayers, facilitating membrane rupture: in artificially stacked double-layer graphene, the grains of one layer are likely to bridge the grain boundaries of the other layer and therefore increase the mechanical stability of the double layer. [53]

The highest yield is seen for the commercial artificially stacked double-layer graphene on Fraunhofer IZM substrates (**Figure S3**), where the spacing between the individual cavities ranged from 3 µm to 13.5 µm. This spacing is larger than the spacing of only 1.5 µm between the cavities on substrates used for the other samples. The larger surface area between the cavities is thought to enhance adhesion and lower the chance of membrane rupture propagation between neighboring cavities. However, the larger spacing decreases the active area of the membrane-based pressure sensors dramatically, reducing their sensitivity, since the unsuspended graphene does not experience strain from a pressure change and therefore does not contribute to a resistance change [13]. An overview of the membrane spacing and the relative membrane area in the different device designs is provided in **Table S2**. Plots of the yield against the relative cavity area are shown in **Figure S4**. As expected, the yield drops with increasing relative cavity area, *i.e.*, with decreasing spacing between cavities. However, conclusions may not be drawn from this evaluation since the relative cavity area correlates with the cavity diameter in the used layouts. To evaluate the effect of the relative cavity area on the yield properly, new substrates with the same cavity diameter but different spacing would be required.

The PMMA removal method seems to have only a minor influence on the yield, especially for double-layer graphene membranes (compare com-G2-A and com-G2-S in **Figure 2**d). For single-layer graphene membranes, PMMA removal by annealing leads to a slightly higher yield than PMMA removal in liquid solvent, up to 1.5 times higher for the smallest membrane diameters, *i.e.*, 1.5 μm. This can be seen when comparing the orange-colored (com-G1-A) and the yellow-colored (com-G1-S) bars in **Figure 2**d.

For some cases, the yield showed significant variation between different samples of the same sample group, as indicated by the error bars. This variation is likely related to the manual nature of the transfer process and could potentially be improved by implementing a transfer process in an automated tool to avoid irregularities and guarantee reproducibility.

The total yield of intact membranes of the graphene from ANL is lower than that of the commercial graphene transferred by our in-house method. It is known that wafer level adhesion affects the graphene layer integrity and subsequent device yield. The wafer level surface structure should thus be optimized for getting a good yield with the dry transfer method from ANL. This optimization was not yet done within this project leading to a lower yield. When disregarding the entirely delaminated pieces of the graphene from ANL, the yield is slightly higher than that on the in-house-grown double-layer graphene samples, up to 1.55 times higher for the 3.4 μm diameter membranes, with larger yield spreads for other diameters. This result shows the potential of the wafer-scale dry transfer method.

Our membrane fabrication process for closed cavities returns higher yields than previously reported in the literature (see **Table S1**). However, it must be noted that the cavity depth plays a significant role, since shallow cavities increase the risk of the graphene membrane adhering to the cavity bottom surface during the transfer process, leading to non-suspended or ruptured membranes. Even though this issue is more severe with wet transfer processes, it also affects dry transfer processes, as a more shallow cavity increases attractive forces (likely dispersive/adsorptive adhesion) between the cavity bottom and the graphene membrane, which may cause stiction of the membrane at the cavity bottom.

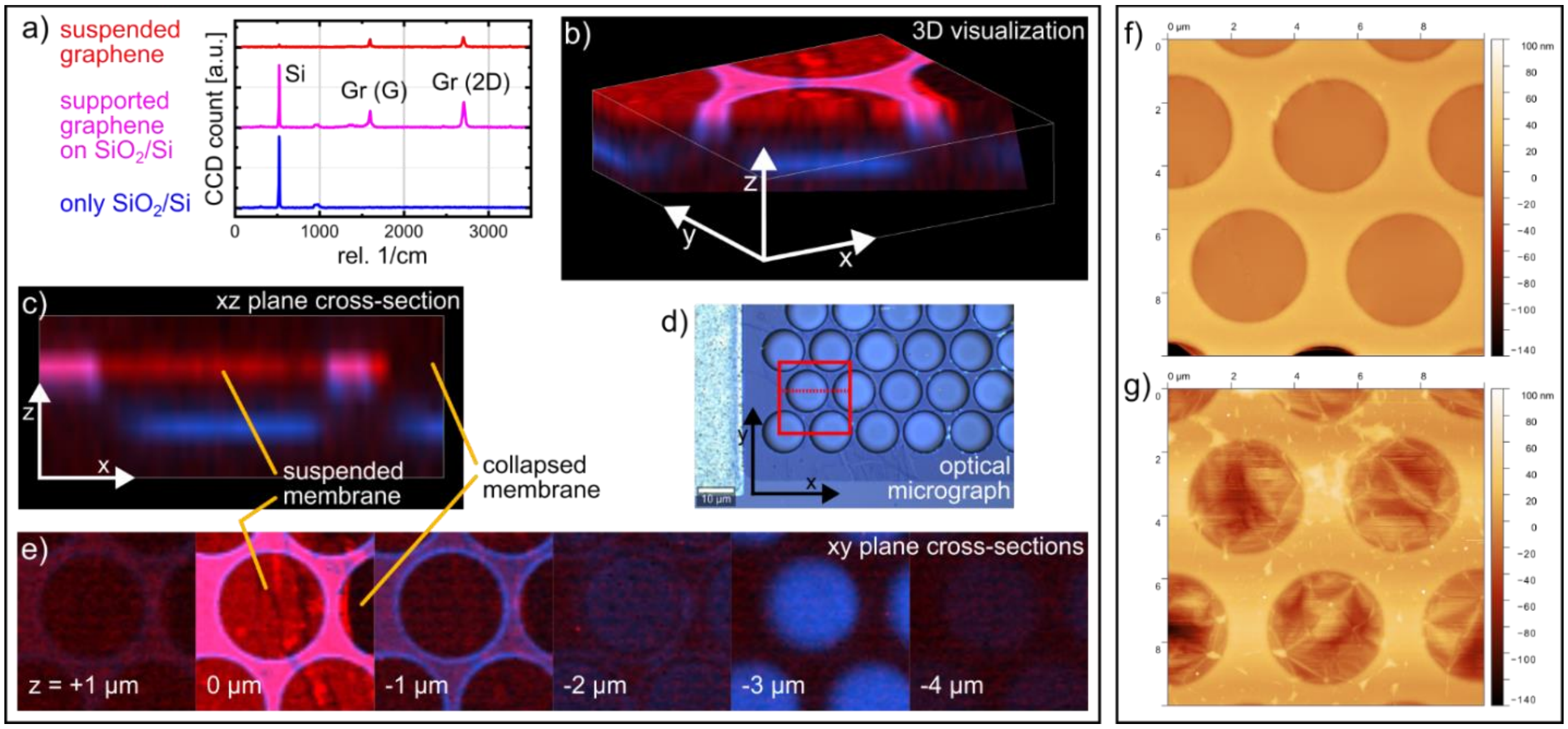

**Figure 3.** (a-e) Raman tomography of suspended graphene membranes: (a) Raman spectra of the three distinguished spectral components. (b) 3D visualization of the Raman tomography. (c) xz-plane cross section along the dotted line shown in (d). (d) Optical micrograph of the scanned area. (e) xy-plane cross sections of the square area shown in (d) at varied focus (z-coordinate). (f-g) Exemplary AFM scans of (f) single-layer and (g) double-layer graphene membranes from commercial graphene.

Raman tomography was performed on some samples (**Figure 3**a-e), confirming that the graphene membranes are truly suspended across the cavities. Three signature spectra were identified (**Figure 3**a): a Si signal (blue), a Si and unsuspended/supported graphene signal (pink), and a suspended graphene signal (red). The areas of dominance of those spectra were then highlighted in the respective color in the volume (**Figure 3**b) and area (**Figure 3**c) cross-sectional images. The individual xy-planes at various focus levels (z-coordinate) are also shown (**Figure 3**e). The reconstructed images show a suspended graphene membrane in the center and a partially collapsed membrane on the right side.

**Figure 3**g-f displays AFM scans of the membranes made of commercial graphene. While the single-layer graphene membranes (**Figure 3**g) have a very flat and smooth appearance, the double-layer graphene (**Figure 3**f) features small particle-like structures and folds or wrinkles above the cavities. Both are likely the result of the manual stacking of two single-layer graphene layers to form the double-layer graphene, leading to small particles or water/air enclosures between the two graphene layers. Nevertheless, the integrity of both the suspended single- and double-layer graphene membranes is confirmed by AFM. Noise in the amplitude error signal, likely due to oscillations of the membrane with the AFM cantilever, were visible on some scans, especially for the graphene sample from ANL (**Figure S5**).

Six-terminal graphene devices with global back-gating though 90 nm of thermally grown $SiO_2$ were fabricated in the vicinity of the membrane devices to perform four-probe field-effect measurements for extracting the sheet resistance and the charge carrier mobility. Note that gating of the suspended graphene is not possible due to the large distance between the graphene and the substrate at the bottom of the cavities, and the corresponding small capacitance. Later, the six-terminal devices were also used for reference measurements inside the pressure chamber set-up. SEM images of the six-terminal devices and the four-probe measurement scheme are shown in **Figure S6**. Plots of the four-probe field-effect mobility and sheet resistance of the graphene layers vs. the back-gate voltage are shown in **Figure S7**. All samples exhibit p-doping, indicated by the shift of the charge neutrality point to positive gate voltages. This is typical for non-encapsulated graphene. The field-effect mobility maximum is approximately 1,500 cm²/Vs for the in-house-grown double-layer graphene devices near the charge neutrality point. The maximum mobility of the commercially transferred double-layer graphene showed slightly higher maximum mobilities up to 2,000 cm²/Vs. The highest mobility values of up to approximately 2,500 cm²/Vs were observed in the double-layer commercial graphene. In general, devices with single-layer graphene (both in-house and commercial) show lower mobilities, although there is significant spread of the data between the different devices. Most devices showed sheet resistance values between 400 and 10,000 Ω/sq near the charge neutrality point. Higher resistance values measured in some devices are likely related to localized cracks or other kinds of defects in the graphene layers.

Electrical measurements of the membrane devices of sample com-G2-A-2d inside the pressure chamber are shown in **Figure 4**. The gas pressure in the chamber was modulated between 200 mbar and 1,500 mbar, while the pressure inside the cavity remained approximately at 1,000 mbar ± 20-150 mbar, depending on the cavity volume change due to membrane displacement, which is most severe for the larger diameter membranes. As a result, the membranes were bent and experienced strain, which led to a change in the sensor resistance. Resistance minima are observed for both the over- and the underpressure case. The resistance maximum is observed when the membranes experience a minimum of strain, here at a pressure of approximately 850 to 950 mbar (indicated with small circles in **Figure 4**a-c). This implies that there is a small under-pressure inside the cavities which might have formed during the transfer process due to the heating. A sensitivity of 0.5 to $3.0 \cdot 10^{-6}$ $mbar^{-1}$ is extracted from 5 pressure sensors with membrane diameters from 3.4 µm to 12 µm. The decreasing resistance for the strained graphene membranes implies a negative piezoresistive GF for the artificially stacked double layer graphene. This was confirmed by our bending beam experiments of the double-layer commercial graphene, which resulted in a small negative GF of approximately -0.46 for tensile strain and -0.61 for compressive strain (**Figure S8**a-b). The sensor response shape is similar to that observed for $PtSe_2$ membrane devices without PMMA as shown in our previous work. [43] The extracted sensitivity values are plotted against the membrane diameter in **Figure 4**d. The devices showed no significant sensitivity increase with increasing membrane diameter, although this would be expected. A possible explanation may by the lower yield of intact membranes at larger diameters, leading to a smaller actual active area (*i.e.*, the area covered with intact membranes per device, calculated as the product of the theoretical suspended membrane area relative to the total channel area, and the yield of intact membranes in the respective device). This, however, could neither be confirmed nor disproven due to the limited amount of data with varying yield for the same membrane diameter and sample. The sensitivity is plotted against the actual active area in **Figure 4**e. In fact, the device with 12 µm membrane diameter possessed similar actual active membrane area as devices with smaller membrane diameter of the same sample, diminishing the expected sensitivity increase from larger membranes.

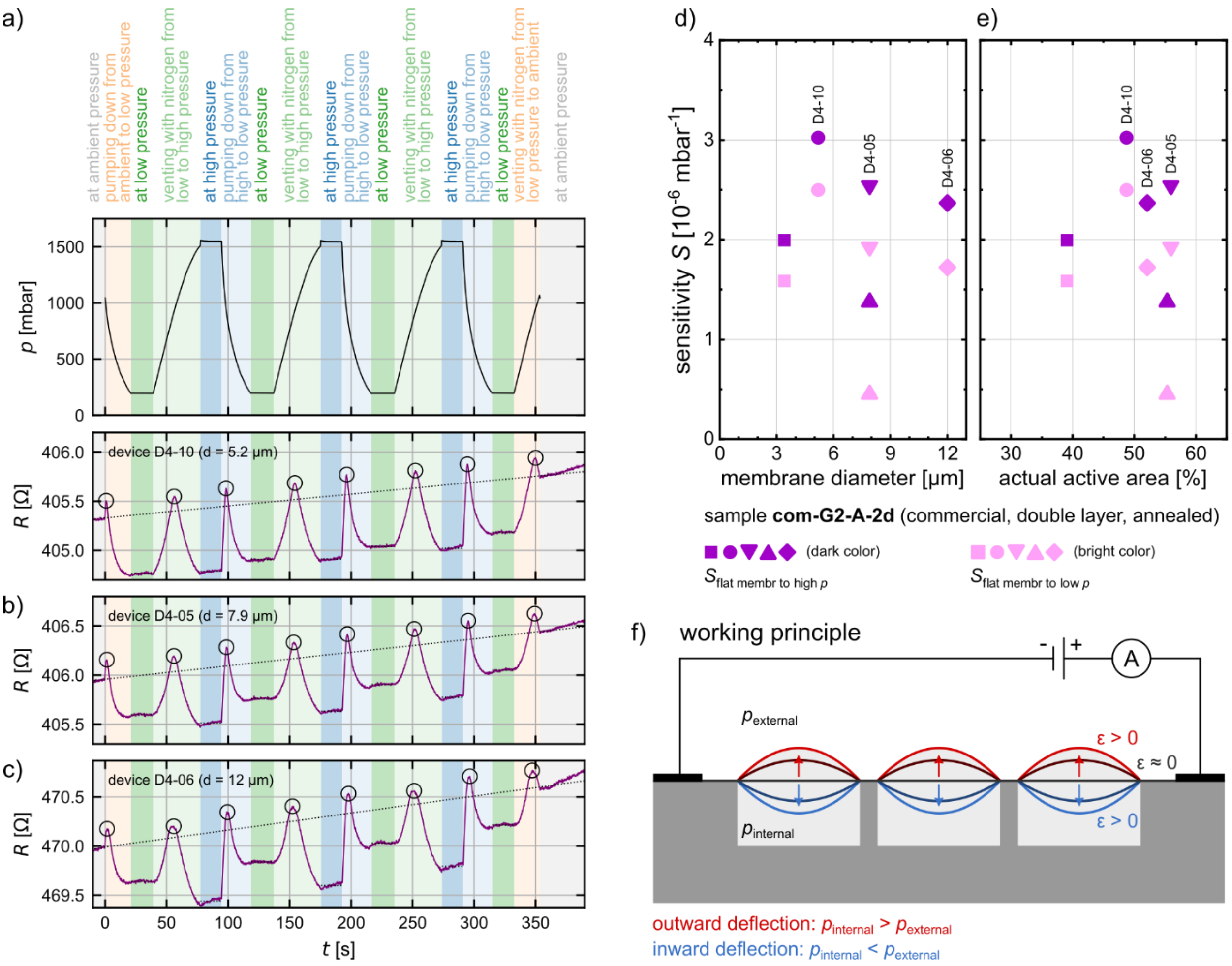

**Figure 4.** Exemplary electrical measurements of graphene pressure sensors of double-layer commercial graphene samples. The measurements of three devices with different membrane diameters are plotted in (a), (b), and (c), as indicated. On all shown devices, the PMMA has been removed by annealing. The pressure is recorded by a reference sensor as shown in the top panel of (a) with the different steps indicated by annotation and background color. The reference pressure curves for (b-c) are not shown but are indicated by the background colors (same levels as in (a)). (d-e) Extracted sensitivity values from the pressure sensor measurements plotted versus membrane diameter and actual active area (*i.e.*, maximum suspended area relative to total channel area, multiplied by yield of intact membranes). The sensitivity was extracted between the resistance maxima

(turning points at flat membrane state, marked by circles in (a)) and the neighboring resistance minima at high and low pressure. (f) Illustration of the working principle.

The observed sensitivity values of sample com-G2-A-2d are in agreement with the previously reported values for piezoresistive graphene membrane pressure sensors, [11,54,55] achieved with high yield, large-scale processes.

Pressure sensor devices on the three additional samples inH-G2-A-2322, inH-G1-A-2332, and ANL-G2-A-3 were also measured in the pressure chamber set-up. However, only a response correlating with the small change in relative humidity (between approximately 5-8 % at low pressure and approximately 9-18 % at high pressure) was observed. This signal matched the response of reference devices without membranes on the same samples (six-terminal devices as shown in **Figure S6**). Graphene is known to be sensitive to changes in relative humidity, [56] which appears to manifest itself here on a low level. Reasons for the absence of a pressure-related response might be small cracks or holes in the graphene membranes (potentially especially in the in-house-grown graphene of quality inferior to the commercial graphene) or insufficient adhesion and sealing (potentially for the ANL samples, where adhesion issues were present), leading to the lack of hermeticity. This insufficient hermeticity remains invisible in the structural characterization. In consequence, the importance of high-quality graphene for the fabrication of truly air-tight membranes must be emphasized.

## Conclusions

We have demonstrated the scalable fabrication of millions of suspended membranes of monolayer and artificially stacked double-layer graphene across closed cavities with a high yield of up to 99 %. A neural-network-based object detection routine for automatically collected SEM images enabled the evaluation of more than 2,000,000 membranes and the extraction of the process yield. The yield is a function of membrane diameter, with the best monolayer graphene samples reaching 85 % for membrane diameters of 5.2 µm and below, and 60 % for a diameter of 7.9 µm. The yield for artificially stacked double-layer graphene membranes was over 95 % for membrane diameters of 7.9 µm and below on the best samples. We have employed the technology for graphene-based NEMS pressure sensors with a sensitivity of $3.0 \cdot 10^{-6}$ $mbar^{-1}$. Our approach is scalable to large-area fabrication, as demonstrated with a wafer-scale, frame-based semi-dry graphene transfer within the 2D-EPL multi-project wafer run, [57] or the proprietary wafer-scale transfer method of ANL in this work. The technique is further transferable to other 2D materials with minor process modifications to exploit the superior piezoresistive properties of those materials, *e.g.*, the transition-metal dichalcogenides $MoS_2$ and $PtSe_2$. [44]

## Methods

### *Substrate fabrication*

Target substrates were fabricated at RWTH/AMO from 150 mm p-doped Si wafers prior to the transfer of the graphene. Stepper lithography was performed for the marker layer and the markers were etched by reactive ion etching (RIE) with a chemistry of $C_4F_8$ and $SF_6$. Another stepper lithography step followed for the cavity layer and 2.0-2.5 µm deep cavities were etched into the Si using RIE with the same $C_4F_8/SF_6$ chemistry. Afterwards, the wafers were cleaned in Piranha solution ($H_2O_2$ and $H_2SO_4$) and thermally oxidized to grow 90 nm $SiO_2$ on top of the Si for passivation. The oxide growth after the etching passivates the cavity sidewalls and floors and avoids short-circuiting from collapsed graphene membranes later. Next, stepper lithography with a resist double layer of lift-off resist (LOR) 3A and the photoactive AZ5214E resist was done, followed by the deposition and patterning of bottom contacts of 5 nm titanium (Ti) and 10 nm nickel (Ni) (both sputtered). Here, the resist stack ensured no metal residues remained after lift-off despite the uneven surface across the etched cavities. A last stepper lithography was done to structure the 300 nm thick aluminum (Al) pads on top of the previously patterned metal contacts in the probing pad areas with a lift-off process for subsequent wire-bonding. Finally, some wafers were diced into $2 \times 2$ cm$^2$ dies.

In parallel, additional 150 mm target substrates were fabricated by Fraunhofer IZM. Cavities of 3-5 µm depth were etched into highly p-doped Si by deep reactive ion etching (DRIE). Then, the substrates were thermally oxidized to grow 100 nm of $SiO_2$. Bottom contacts of 5 nm Ti and 15 nm palladium (Pd) were evaporated and patterned by lift-off. Afterward, probing pads of titanium tungsten (TiW) and gold (Au) were deposited by sputtering and electroplating and patterned by wet chemical etching. The results retrieved from these wafers are included to demonstrate the process feasibility on wafer-scale substrates.

***Graphene preparation and transfer***

Different types of graphene were used for the suspended membranes. Commercially available single-layer CVD graphene on copper (Cu) foil (Grolltex, Inc.), as well as CVD-grown single-layer graphene on Cu foil grown at 1060 °C in an in-house inductively-coupled plasma chemical vapor deposition (ICP-CVD) tool (Oxford Instruments), were employed for the transfer of graphene onto 2 × 2 cm² dies. Both types of graphene were coherent and largely monolayer.

Graphene from the two sources was also artificially stacked to form double layer graphene before the transfer onto the target substrates by wet-transferring one layer of graphene with supporting PMMA A6 950k (thickness: 550 nm) onto another Cu foil with graphene. [45–48] Such artificially stacked double-layer graphene was also transferred onto the two 6’’ wafers from Fraunhofer IZM.

To create the artificially stacked double-layer graphene, PMMA A6 950k was spin-coated onto one Cu foil with monolayer graphene (approximately 10 cm × 6 cm), attached to a Si dummy wafer for handling. After subsequent baking at 120 °C for 10 minutes, the Cu foil was removed from the Si wafer, turned upside down, and carefully taped to a glass carrier. A low-power oxygen plasma was used to remove any back-side graphene from the Cu foil. Afterwards, the Cu foil was removed from the glass carrier. To etch the Cu, a solution of $H_2O_2$, $H_2SO_4$, and $H_2O$ was prepared, and the Cu foil was placed on the surface of the solution with the graphene/PMMA side facing up. After a few minutes, involving rinsing steps in de-ionized (DI) water and changes of the etching solution, the Cu foil was entirely etched, resulting in only a film of graphene covered with PMMA floating on the surface of the solution. The film was transferred from the etching solution to a beaker with only DI water by means of “fishing” with a plastic card. Finally, a second piece of graphene on Cu foil, slightly larger than the first piece, was attached to the plastic card and positioned under the film of graphene/PMMA in the beaker, to then attach the two graphene films to each other by “fishing” the graphene/PMMA film with the Cu foil/graphene from below. The resulting stack of Cu foil, graphene, graphene, and PMMA was left to dry in air for several hours before cutting into smaller pieces for individual transfer with transfer frames.

The transfer onto the target substrates was done using a hot and dry frame-assisted transfer process that facilitates the handling, rinsing, and drying of the graphene/PMMA films. [44]

PMMA A6 950k was applied by spin-coating (1 minute at 3000 rpm) onto the graphene on Cu foil attached to a Si dummy wafer for handling. After spin-coating, the PMMA was baked on the stack of Si dummy wafer, Cu foil, and graphene for 10 minutes at 120 °C. Square pieces of approximately 2-3 cm edge length of single or double-layer graphene on Cu foil covered with PMMA were then attached to frames of transparent printer foil (number 3560 from Avery Zweckform, polyester, 0.1 mm thickness) utilizing heat-resistant polyimide tape (Kapton, number 5413 from 3M). The Cu foils were entirely removed by wet etching in diluted HCl and $H_2O_2$, leaving only the frames with graphene/PMMA films. After rinsing the graphene/PMMA frames in DI water, the frames were removed and dried in air. The target substrates were placed on top of a 115 °C hotplate and the dried graphene/PMMA frames were carefully placed on top of the target substrates. The hotplate temperature was gradually increased up to 180 °C to promote the adhesion of the graphene to the substrates. The heat softens the PMMA film, enabling the graphene to adhere smoothly to the target substrates without significant wrinkles or bubbles. The frame was subsequently removed by cutting the graphene/PMMA with a sharp scalpel, leaving the graphene/PMMA on the target substrate.

In addition to the in-house dry transfer process, graphene growth and its transfer onto a prepared 150 mm cavity substrate was procured as a commercial service from ANL. A 4'' piece of double layer CVD graphene was grown, artificially stacked, and then dry-transferred with PMMA using a proprietary method. The wafer was then carefully cleaved into 2 × 2 $cm^2$ dies for the following graphene patterning process.

***Graphene patterning and PMMA removal***

Graphene/PMMA was patterned into devices on chip-scale using contact lithography. The PMMA was not removed before the patterning but acted as a mechanic support layer for the suspended graphene during the fabrication process. To prevent the mixing of the photoactive AZ5214E resist and the PMMA, which form a generally insoluble layer at their interface, LOR 3A was first deposited on the PMMA, resulting in a layer stack of single/double layer graphene, PMMA, LOR 3A, and AZ5214E. A TMAH-based developer was used after exposure to remove both the AZ5214E resist and the LOR in the exposed areas. The PMMA and graphene were then etched by a low-power $O_2$ plasma in RIE. After etching, the AZ5214E photoresist and LOR were removed by a flood exposure and a subsequent 30 s step in TMAH-based developer. A microscope image of a fabricated device prior to PMMA removal with its device dimensions indicated is shown in **Figure S9**. The PMMA was finally removed either in warm acetone (30 min at 60 °C), careful rinsing in isopropyl alcohol (IPA) and natural drying of the IPA in air, or by annealing in inert atmosphere at 450 °C, resulting in suspended single and double layer graphene membranes.

***Scanning electron microscopy analysis and yield evaluation***

The samples were examined in an SEM at a tilt angle of 20 ° to evaluate the yield of intact suspended graphene membranes after the fabrication process. The SEM visualizes the cavity sidewalls and creates a realistic three-dimensional image of the structure. A macro script was used to collect hundreds of high-resolution images of the various samples. Each sample contained multiple instances of six different devices with cavity diameters ranging from 1.5 µm to 12 µm. Devices with cavity diameters of 1.5 µm, 2.3 µm, and 3.4 µm were imaged at 1,200 X magnification and in four segments, while devices with cavity diameters of 5.2 µm, 7.9 µm, and 12 µm were imaged at 640 X magnification in single images. This resulted in 120 images of the 48 devices of each of the 9 segments on each die. In total, more than 1,600 high-resolution (3,072 px × 2,304 px) SEM images were collected across the different samples.

The SEM images were then processed by a Python script employing the open-source software library TensorFlow for machine learning and artificial intelligence. The TensorFlow model EfficientDet D2 [49] for object detection in images was trained first with manually labelled images and then with automatically labelled images that were manually corrected until a sufficient detection precision of intact and broken membranes was achieved. Membranes were considered broken as soon as small holes were visible in the graphene, even though most broken membranes were easily identifiable due to entirely collapsed graphene and the bright appearance of the cavity sidewalls. Each of the collected SEM images was cropped into 16 segments to decrease the image file size for faster computation. Additionally, brightness and contrast levels were adjusted. Next, the images were analyzed with the previously trained model to recognize and count both the intact, suspended and the broken, collapsed graphene membranes.

***Raman tomography***

Raman cross-sectional depth scanning (Raman tomography) [50] was performed using a WITec alpha300 R Raman microscope set-up with a 532 nm laser. The focal point of the laser beam had an approximate diameter of 300 nm and a depth of ≤ 1 µm in the direction of the beam. The lateral pixel size was therefore chosen as ⅓ µm while the pixel spacing in z-direction was set to 1 µm.

Several Raman scans at varied focus levels allow rendering a three-dimensional (3D) cross-sectional image of the cavities and the suspended graphene, created using the Volume Viewer plugin of the Fiji software. [51]

***Atomic force microscopy***

AFM was conducted using a Bruker Dimension Icon atomic force microscope in tapping mode. AFM images of the suspended graphene membranes were recorded using a scan rate of 0.3 Hz.

***Electrical characterization***

Electrical measurements of back-gated test structures were performed using a Cascade Summit 12000 A semi-automatic prober connected to a Hewlett Packard 4156B Precision Semiconductor Parameter Analyzer and a Hewlett Packard E5250A Low Leakage Switch Mainframe, and executed by Keysight WaferPro Express test routine software. The field-effect mobility was extracted from four-probe field-effect measurements on six-port devices (**Figure S6**) according to $\mu_{\mathrm{F}} = (\partial(I_{\mathrm{D}}/V_{\mathrm{diff}})/\partial V_{\mathrm{BG}}) \cdot L_{\mathrm{inner}} \cdot d_{\mathrm{ox}}/(W \cdot \varepsilon_{\mathrm{r,ox}} \cdot \varepsilon_0)$, where $I_{\mathrm{D}}$ is the current along the graphene channel, $V_{\mathrm{diff}}$ is the voltage between the two inner contacts, $V_{\mathrm{BG}}$ is the back-gate voltage, $L_{\mathrm{inner}} \approx 25$ µm is the distance between the two inner contacts, $d_{\mathrm{ox}} = 90$ nm is the thickness of the gate oxide, $W \approx 10$ µm is the graphene channel width, and $\varepsilon_{\mathrm{r,ox}} = 3.9$ and $\varepsilon_0$ are the relative permittivity of the gate oxide and the vacuum permittivity, respectively. The precise values for $W$ and $L_{\mathrm{inner}}$ were determined from SEM images for each sample. The sheet resistance was extracted from the same measurements using $R_{\mathrm{sheet}} = V_{\mathrm{diff}} \cdot W/(I_{\mathrm{D}} \cdot L_{\mathrm{inner}})$.

Some samples were further diced into 6 × 6 $mm^2$ chips and then wire-bonded into 44-pin ceramic chip carriers (LCC44) using a 25 µm diameter Au wire with a tpt HB16 wire bonder by ball-bonding (**Figure S10**). The chip carriers were then inserted into a pressure chamber built in house, [52] but modified to enable automation of the valves for pressure control and to facilitate synchronized data acquisition of the device under test (DUT) with included reference sensors (**Figure S11**). The chamber pressure was repeatedly pumped down to 200 mbar and subsequently increased up to 1,500 mbar using nitrogen ($N_2$) gas. A Keithley 4200-SCS parameter analyzer recorded the electrical signals of the graphene-membrane-based pressure sensors. The sensitivity $S$ was extracted according to $S = \Delta R/(R_0 \cdot \Delta p)$, where $R_0$ is the device resistance at ambient pressure $p_0$, and $\Delta R = R - R_0$ is the difference in resistance at the applied pressure difference $\Delta p = p - p_0$.

**Supporting Information.** Detailed information on intact membrane yield and dimensions, including comparisons with literature; additional SEM images of graphene membranes, additional AFM scan, SEM images of six-port devices, electrical measurement data, piezoresistive gauge factor measurement data, photographs of wire-bonded sample and pressure chamber set-up.

**Corresponding Author**

* Max C. Lemme (max.lemme@eld.rwth-aachen.de)

**Author Contributions**

The manuscript was written through contributions of all authors. All authors have given approval to the final version of the manuscript.

**Acknowledgements**

This work has received funding from the German Ministry of Education and Research (BMBF) under grant agreements 16ES1121 (ForMikro-NobleNEMS) and 03XP0210 (GIMMIK), from the European Union's Horizon 2020 research and innovation programme under grant agreement 881603 (Graphene Flagship Core 3), and the German Research Foundation (DFG) under grant agreements LE 2441/11-1 (2D-NEMS) and INST 221/96-1. The authors acknowledge the assistance of Martin Otto (AMO GmbH) in preparing double-layer graphene.

# Supporting Information

# High-yield large-scale suspended graphene membranes over closed cavities for sensor applications

*Sebastian Lukas[1], Ardeshir Esteki[1], Nico Rademacher[2], Vikas Jangra[1], Michael Gross[1], Zhenxing Wang[2], Ha-Duong Ngo[3], Manuel Bäuscher[4], Piotr Mackowiak[4], Katrin Höppner[4], Dominique J. Wehenkel[5], Richard van Rijn[5], Max C. Lemme[1,2]*

[1]Chair of Electronic Devices, RWTH Aachen University, Otto-Blumenthal-Str. 25, 52074 Aachen, Germany

[2]AMO GmbH, Advanced Microelectronic Center Aachen, Otto-Blumenthal-Str. 25, 52074 Aachen, Germany

[3]University of Applied Sciences Berlin, Wilhelminenhofstr. 75A (C 525), 12459 Berlin, Germany

[4]Fraunhofer IZM, Gustav-Meyer-Allee 25, 13355 Berlin, Germany

[5]Applied Nanolayers B.V., Feldmannweg 17, 2628 CT Delft, The Netherlands

**Table S1.** Comparison of the highest reported yields of intact CVD graphene membranes on closed cavities.

<table>
<tr><th>material</th><th>transfer method</th><th>cavity depth</th><th>cavity di-ameter</th><th>yield</th><th>error bar</th><th>number of evaluated membranes</th><th>ref.</th></tr>
<tr><td rowspan="8">monolayer CVD gra-phene</td><td rowspan="8">dry, PDMS frame, hot transfer (> 12h at 150 °C), thermal PMMA removal</td><td rowspan="4">2 µm</td><td>2.7 µm</td><td>58 %</td><td>+/- 5 % (approx.)</td><td rowspan="8">unknown</td><td rowspan="8">Suk et al. [1]</td></tr>
<tr><td>3.9 µm</td><td>38 %</td><td>+/- 5 % (approx.)</td></tr>
<tr><td>5.1 µm</td><td>26 %</td><td>+/- 5 % (approx.)</td></tr>
<tr><td>7.3 µm</td><td>13 %</td><td>+/- 5 % (approx.)</td></tr>
<tr><td rowspan="4">300 nm</td><td>2.7 µm</td><td>32 %</td><td>+/- 10 % (approx.)</td></tr>
<tr><td>3.9 µm</td><td>23 %</td><td>+/- 6 % (approx.)</td></tr>
<tr><td>5.1 µm</td><td>13 %</td><td>+/- 2 % (approx.)</td></tr>
<tr><td>7.3 µm</td><td>6 %</td><td>+/- 4 % (approx.)</td></tr>
<tr><td>monolayer CVD gra-phene</td><td>wet, transfer of Al/photore-sist/graphene/Cu stack, dry spinning, metal contacts after transfer, Al wet etching, criti-cal point drying for photore-sist removal, annealing for residue removal (patent pend-ing)</td><td>1 µm</td><td>6 µm</td><td>55.73 %</td><td>+/- 2.62 %</td><td>720 membranes<br>3 samples</td><td>Gupta et al. [2]</td></tr>
<tr><td rowspan="5">monolayer CVD gra-phene</td><td>wet</td><td rowspan="10">1.4 µm</td><td rowspan="10">3 µm</td><td>0 %</td><td>unknown</td><td>≥ 120 membranes</td><td rowspan="10">Wagner et al. [3] and Wagner (PhD thesis) [4]</td></tr>
<tr><td>dry, thermal release tape</td><td>0 %</td><td>unknown</td><td>≥ 120 membranes</td></tr>
<tr><td>dry, PDMS stamp</td><td>2.7 %</td><td>unknown</td><td>≥ 120 membranes</td></tr>
<tr><td>dry, Graphenea proprietary</td><td>20 %</td><td>unknown</td><td>≥ 120 membranes</td></tr>
<tr><td>dry, vacuum-assisted, PDMS</td><td>48 %</td><td>unknown</td><td>≥ 120 membranes</td></tr>
<tr><td rowspan="5">double layer CVD gra-phene</td><td>wet</td><td>16 %</td><td>unknown</td><td>≥ 120 membranes</td></tr>
<tr><td>dry, thermal release tape</td><td>5 %</td><td>unknown</td><td>≥ 120 membranes</td></tr>
<tr><td>dry, PDMS stamp</td><td>0 %</td><td>unknown</td><td>≥ 120 membranes</td></tr>
<tr><td>dry, Graphenea proprietary</td><td>64 %</td><td>unknown</td><td>≥ 120 membranes</td></tr>
<tr><td>dry, vacuum-assisted, PDMS</td><td>72 %</td><td>unknown</td><td>≥ 120 membranes</td></tr>
</table>

(table continued on next page)

| material | transfer method | cavity depth | cavity diameter | yield | error bar | number of evaluated membranes | ref. |
|---|---|---|---|---|---|---|---|
| monolayer CVD graphene | dry, hot, frame-based | 2.0-2.5 µm | 1.5 µm | 87.0 % | max. 96.5 % (+ 9.5 %) min. 74.7 % (- 12.3 %) | 574382 membranes 6 samples (com-G1-A) | this work |
| | | | 2.3 µm | 84.9 % | max. 96.6 % (+ 11.7 %) min. 74.2 % (- 10.7 %) | 379337 membranes 6 samples (com-G1-A) | |
| | | | 3.4 µm | 78.6 % | max. 93.4 % (+ 14.8 %) min. 57.1 % (- 21.5 %) | 216300 membranes 6 samples (com-G1-A) | |
| | | | 5.2 µm | 57.0 % | max. 88.7 % (+ 31.7 %) min. 30.1 % (- 26.9 %) | 113217 membranes 6 samples (com-G1-A) | |
| | | | 7.9 µm | 22.8 % | max. 60.5 % (+ 37.7 %) min. 4.6 % (- 18.2 %) | 57717 membranes 6 samples (com-G1-A) | |
| | | | 12 µm | 4.0 % | max. 11.5 % (+ 7.5 %) min. 0.3 % (- 3.7 %) | 26866 membranes 6 samples (com-G1-A) | |
| double layer CVD graphene | | | 1.5 µm | 99.8 % | +/- 0.0005 % | 156240 membranes 2 samples (com-G2-A-IZM) | |
| | | | 2.0 µm | 99.0 % | +/- 0.007 % | 156841 membranes 2 samples (com-G2-A-IZM) | |
| | | | 2.3 µm | 85.6 % | max. 99.3 % (+ 13.7 %) min. 75.4 % (- 10.2 %) | 219649 membranes 3 samples (com-G2-A) | |
| | | | 3.0 µm | 98.0 % | +/- 0.007 % | 11599 membranes 2 samples (com-G2-A-IZM) | |
| | | | 3.4 µm | 86.0 % | max. 98.5 % (+ 12.5 %) min. 76.9 % (-9.1 %) | 126555 membranes 3 samples (com-G2-A) | |
| | | | 5.2 µm | 65.1 % | max. 77.6 % (+ 12.5 %) min. 40.9 % (- 24.2 %) | 65737 membranes 3 samples (com-G2-A) | |
| | | | 7.9 µm | 57.0 % | max. 95.3 % (+ 38.3 %) min. 37.6 % (- 19.4 %) | 33801 membranes 3 samples (com-G2-A) | |
| | | | 12 µm | 42.9 % | max. 81.9 % (+ 39.0 %) min. 18.6 % (- 24.3 %) | 15511 membranes 3 samples (com-G2-A) | |

**Table S2.** Membrane diameters and spacings for the different device layouts on the two different substrate types used within the study.

| substrate type | device number | membrane diameter | membrane spacing | suspended graphene (in case of 100 % yield) | |
|---|---|---|---|---|---|
| | | | | absolute within (310 µm)² | relative |
| in-house-fabricated | 01 | 1.5 µm | 1.5 µm | 21659.9 µm² | 22.5 % |
| | 02 | 2.3 µm | 1.5 µm | 32365.6 µm² | 33.7 % |
| | 03 | 3.4 µm | 1.5 µm | 40529.6 µm² | 42.2 % |
| | 04 | 5.2 µm | 1.5 µm | 51224.0 µm² | 53.3 % |
| | 05 | 7.9 µm | 1.5 µm | 60535.6 µm² | 63.0 % |
| | 06 | 12 µm | 1.5 µm | 66161.9 µm² | 68.8 % |
| IZM | V1.5P05 | 1.5 µm | 3.5 µm | 7888.5 µm² | 8.2 % |
| | V1.5P15 | 1.5 µm | 13.5 µm | 890.6 µm² | 0.9 % |
| | V2P05 | 2.0 µm | 3 µm | 14024.1 µm² | 14.6 % |
| | V2P15 | 2.0 µm | 13 µm | 1583.4 µm² | 1.6 % |
| | V3P15 | 3.0 µm | 12 µm | 3562.6 µm² | 3.7 % |

com-G2-A-IZM5

com-G2-A-IZM6

com-G2-A-IZM6

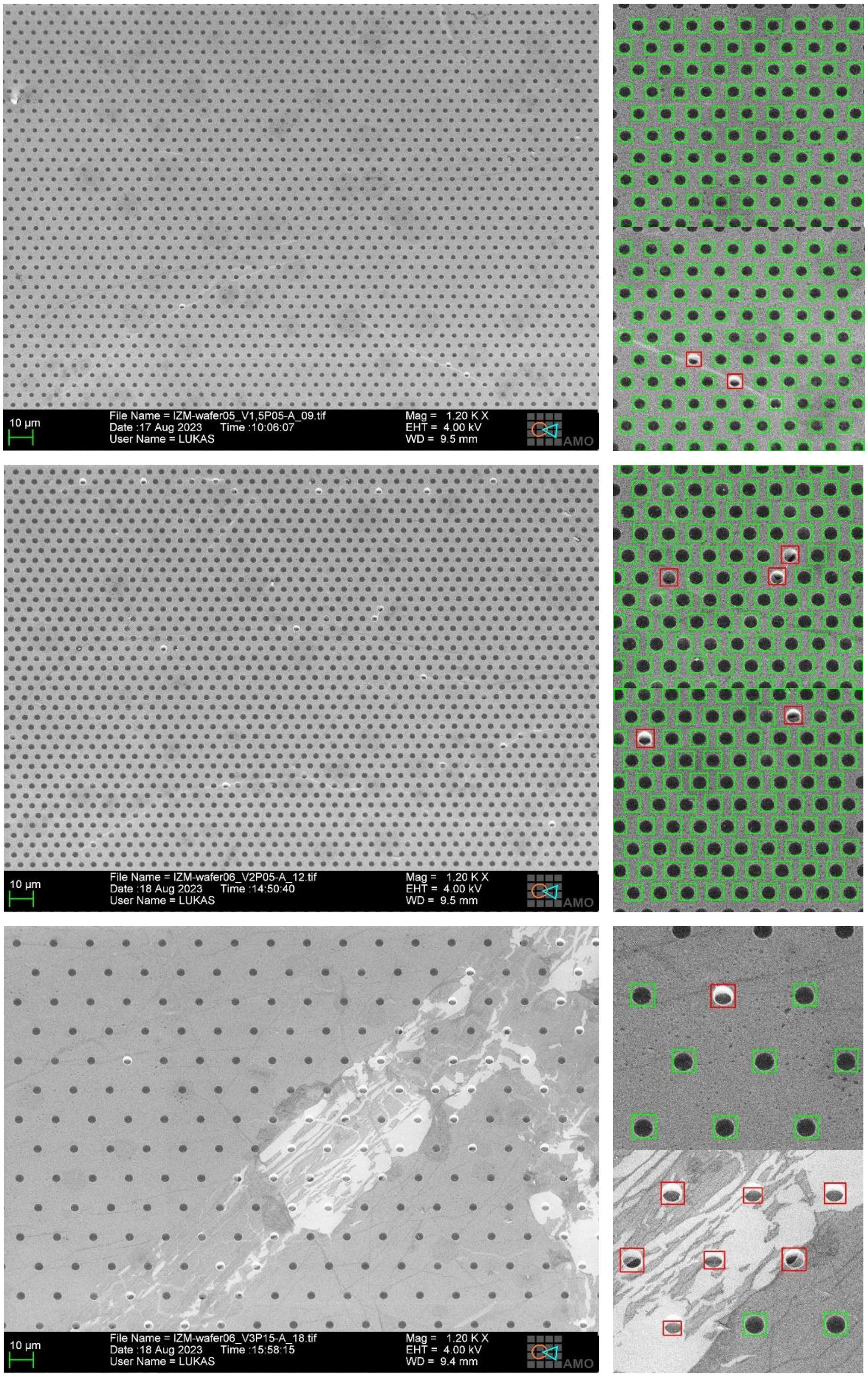

**Figure S1.** More exemplary SEM images of the evaluated graphene membrane arrays before and after yield analysis. The original SEM images are shown on the left, while two cropped images (each 1/16 of original image) after contrast/brightness adjustment and membrane detection are shown to the right of the respective original image. Continued on next page.

com-G2-A-2d

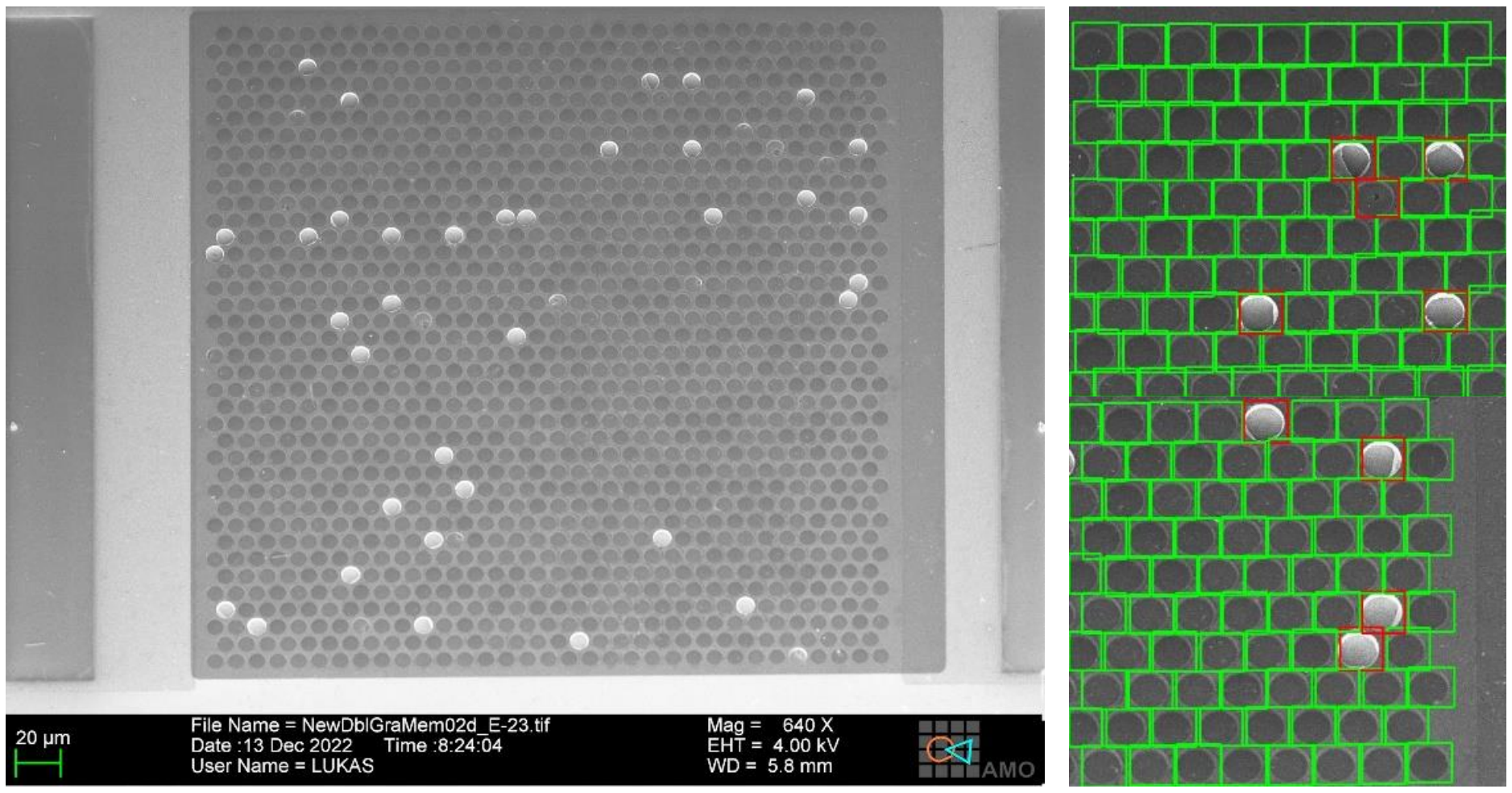

com-G2-S-2a

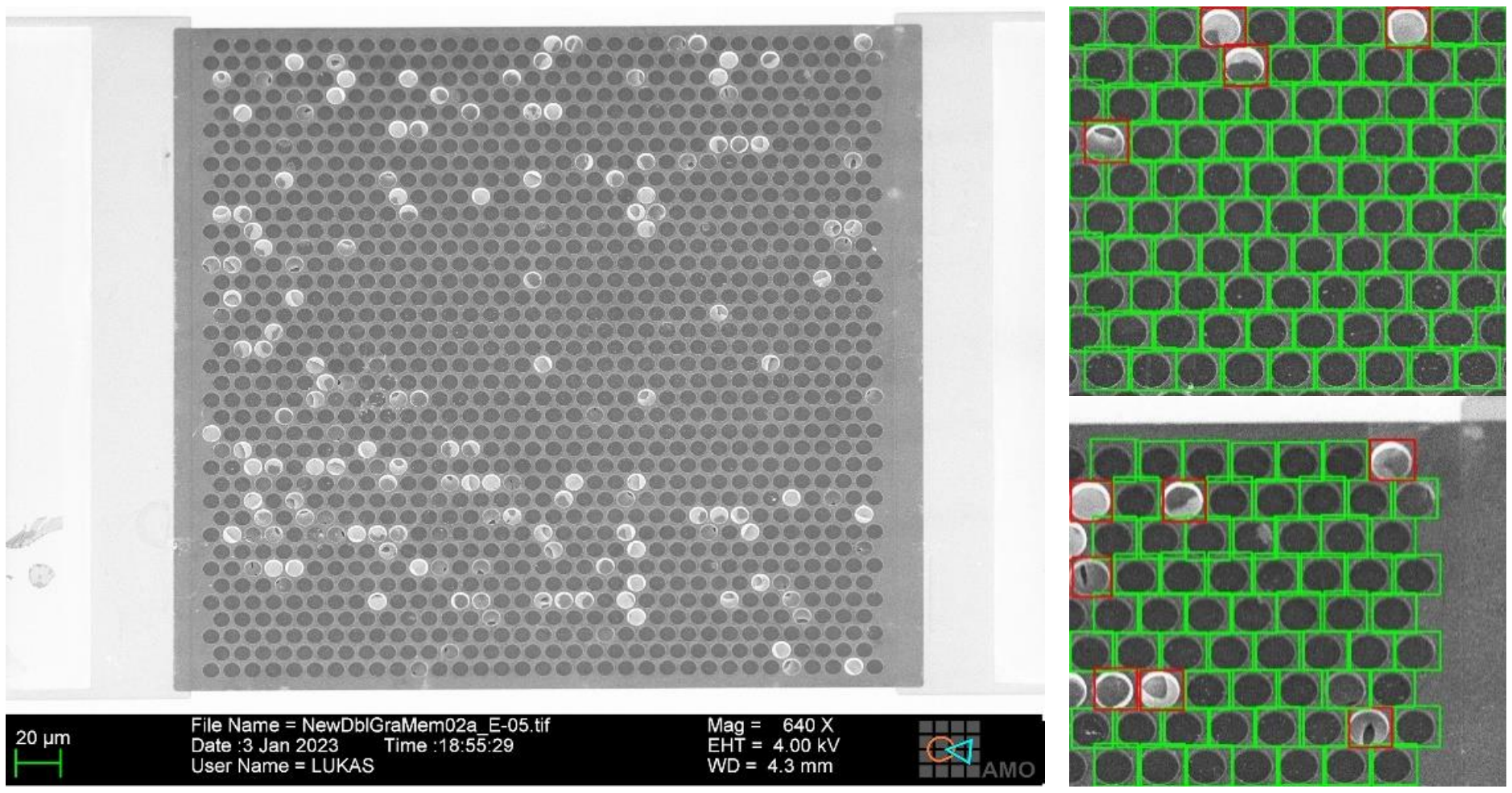

com-G1-A-1b

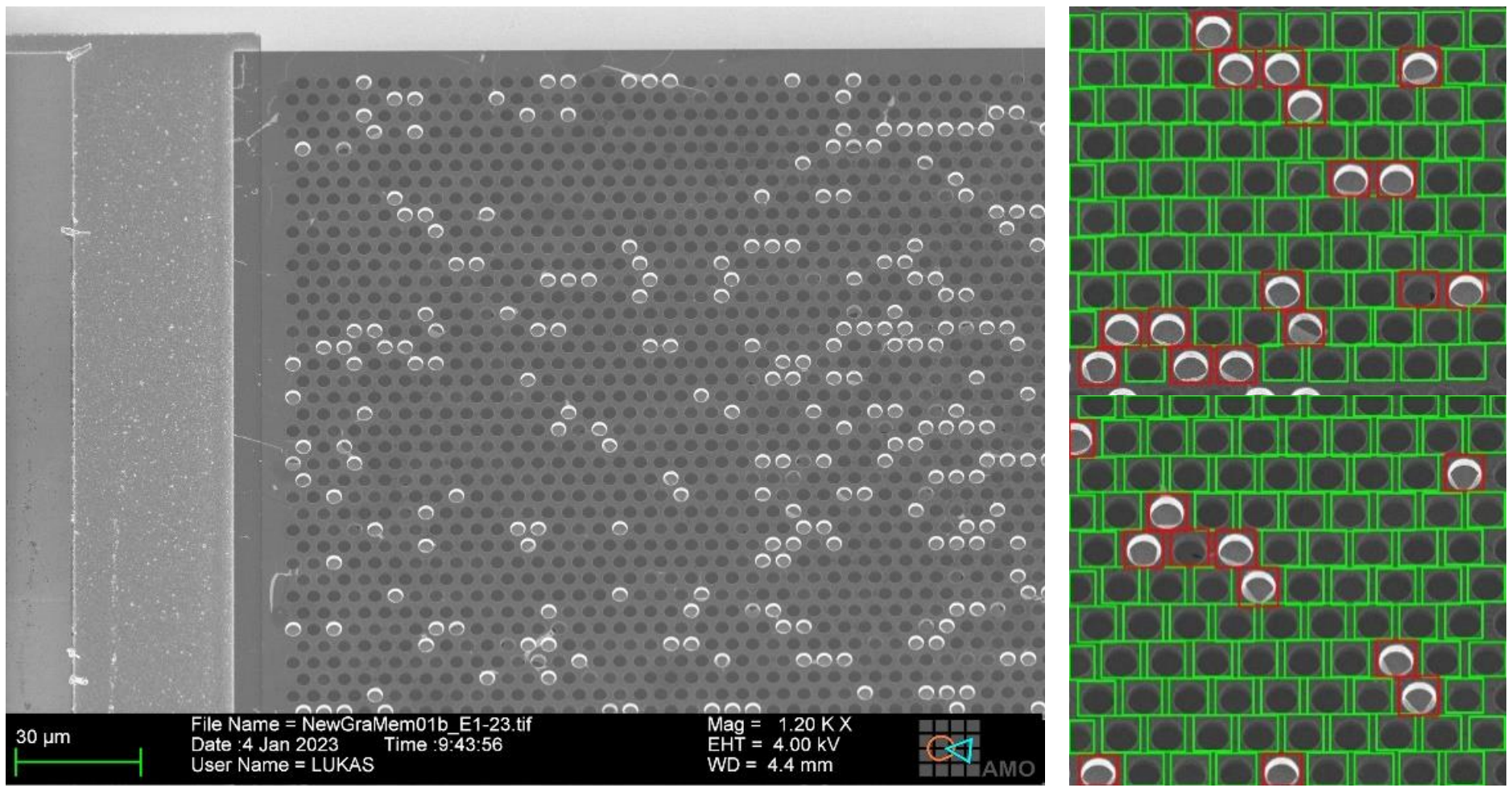

**Figure S1** (continued).

com-G1-S-2a

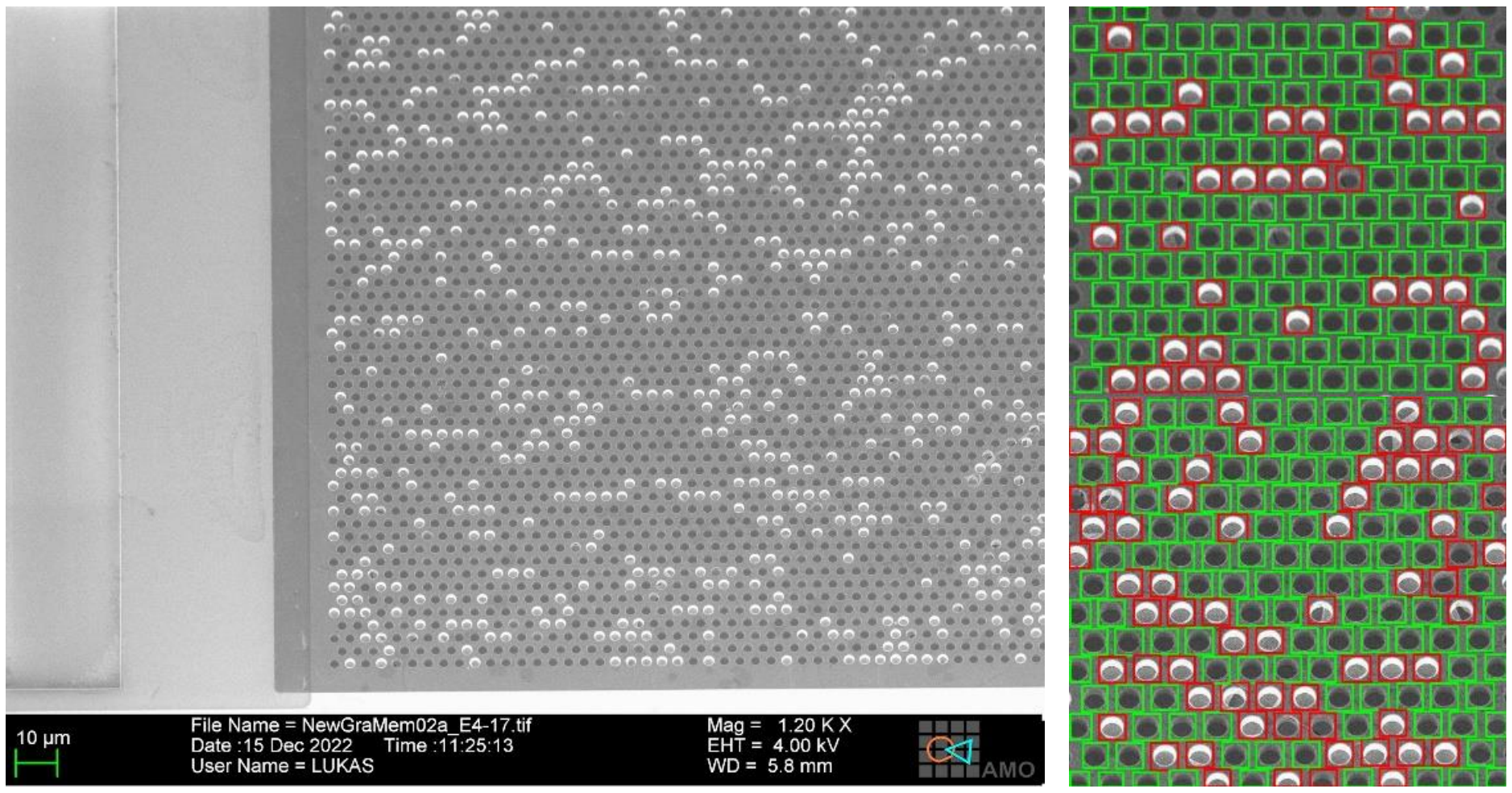

inH-G2-A-2321

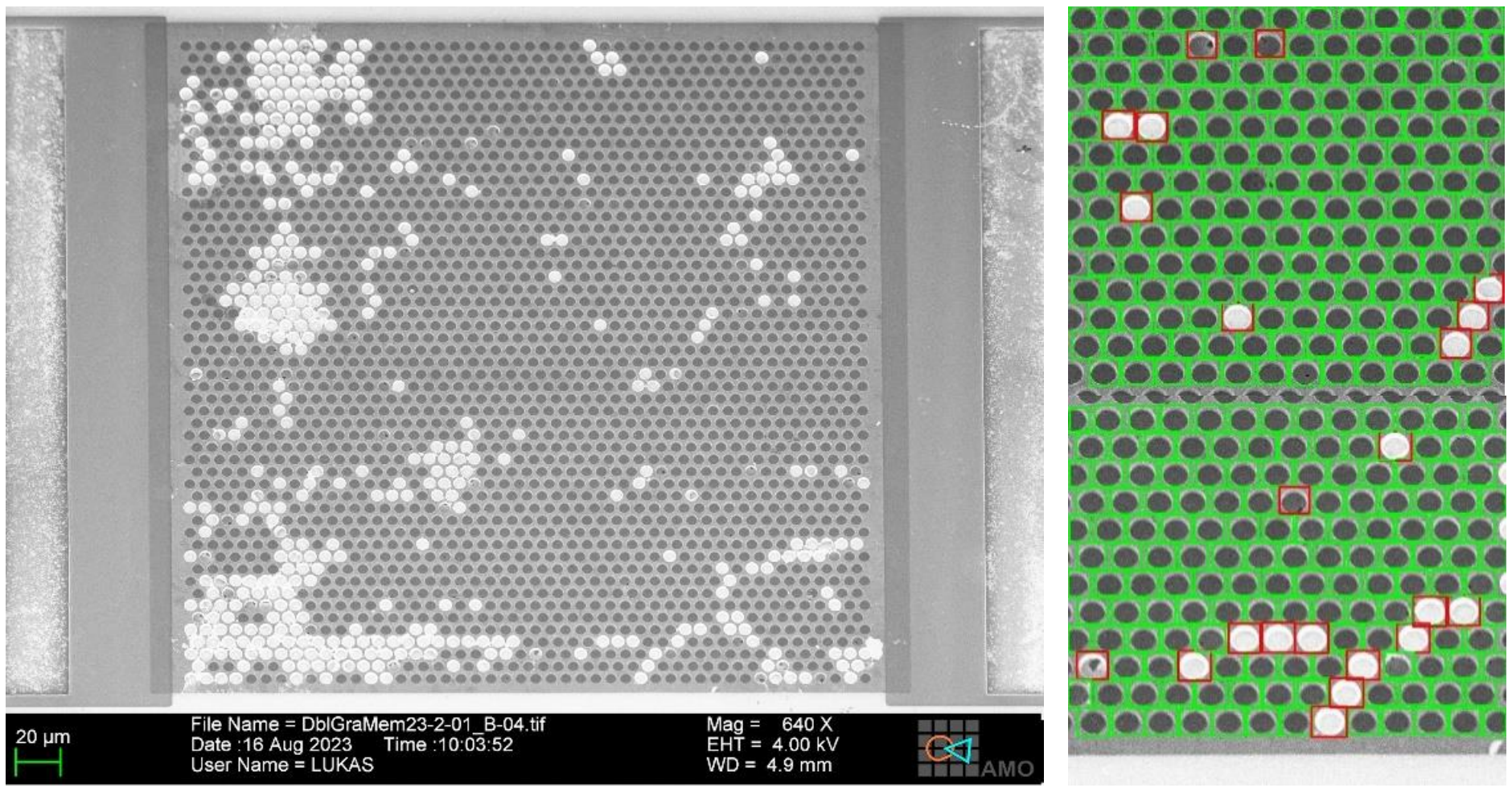

inH-G1-A-2332

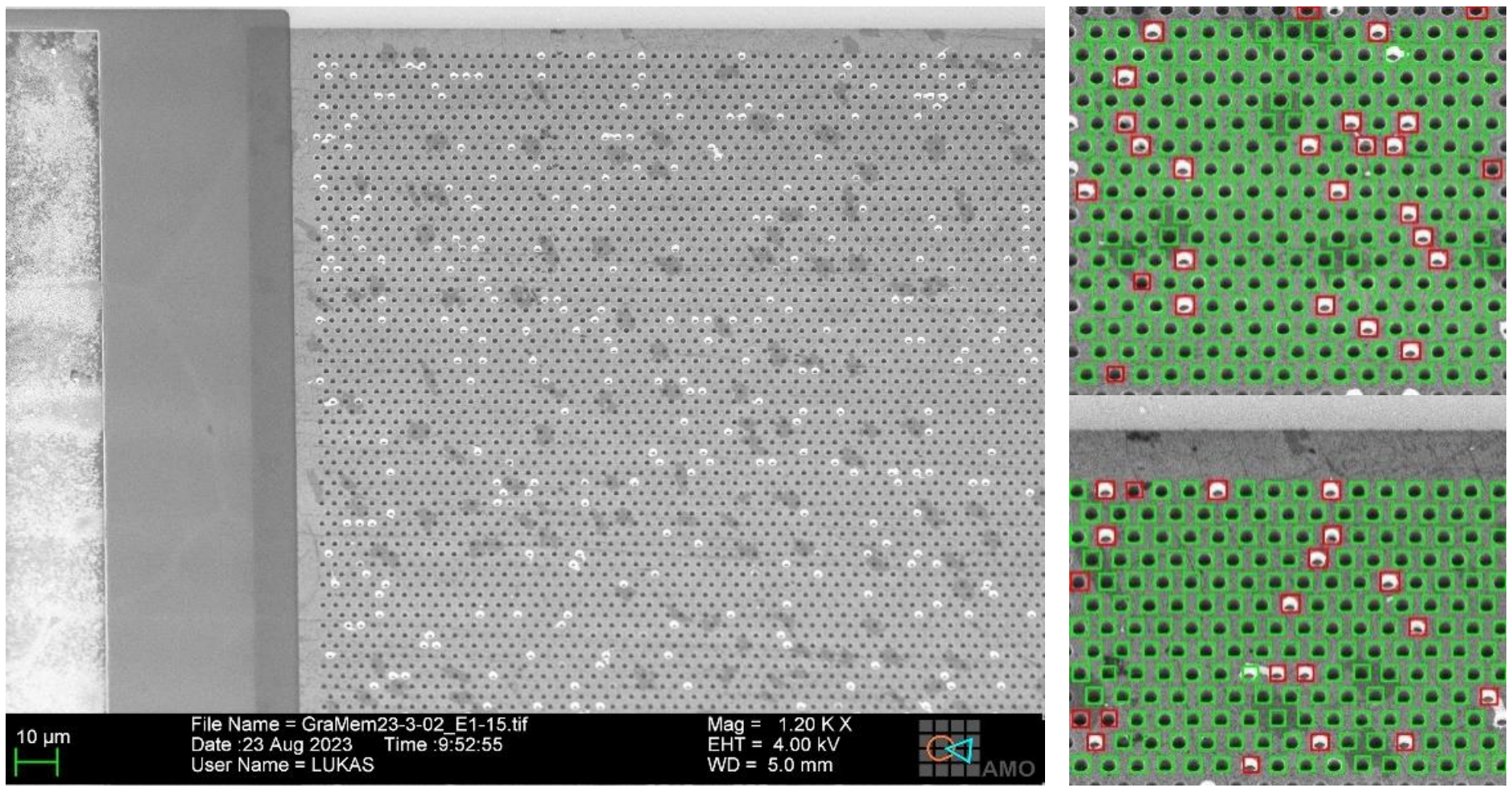

**Figure S1** (continued).

ANL-G2-A-2

ANL-G2-A-3

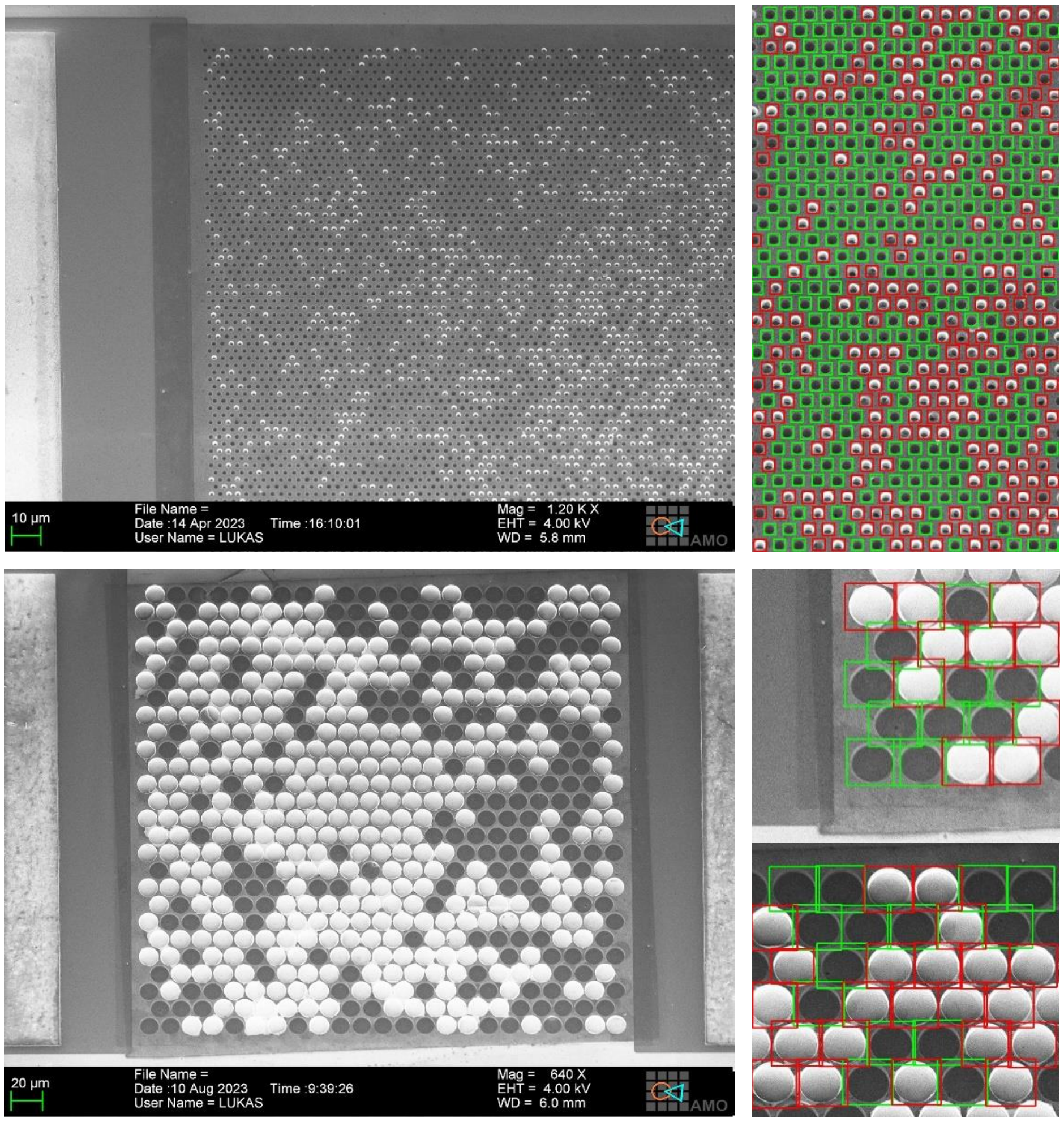

**Figure S1** (continued).

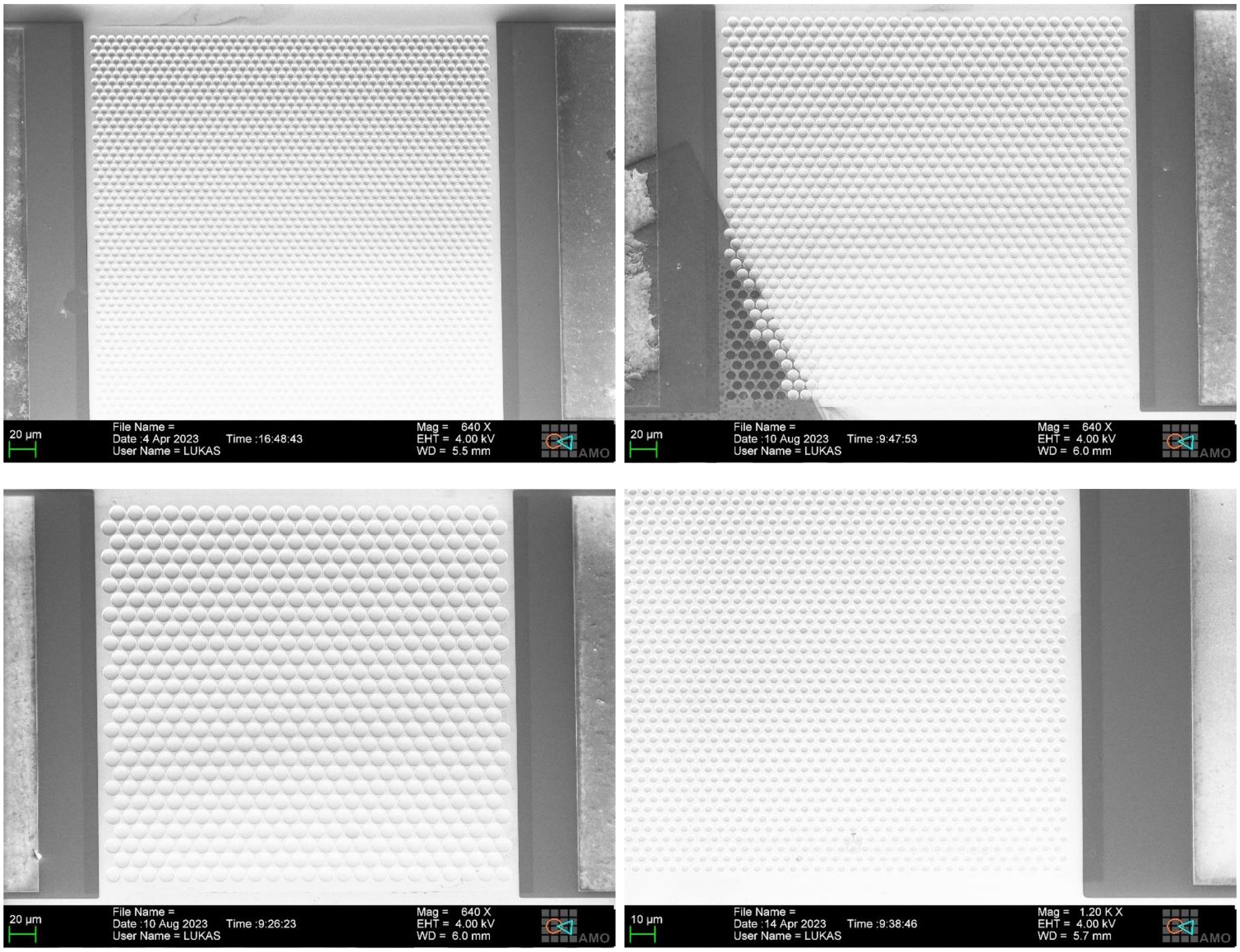

**Figure S2.** SEM images of samples after graphene transfer by ANL and patterning. The graphene delaminated from the devices entirely and washed away. In the top right image, the rectangular patterned graphene patch was redeposited in the bottom left of the device, covering a small fraction of the membrane array.

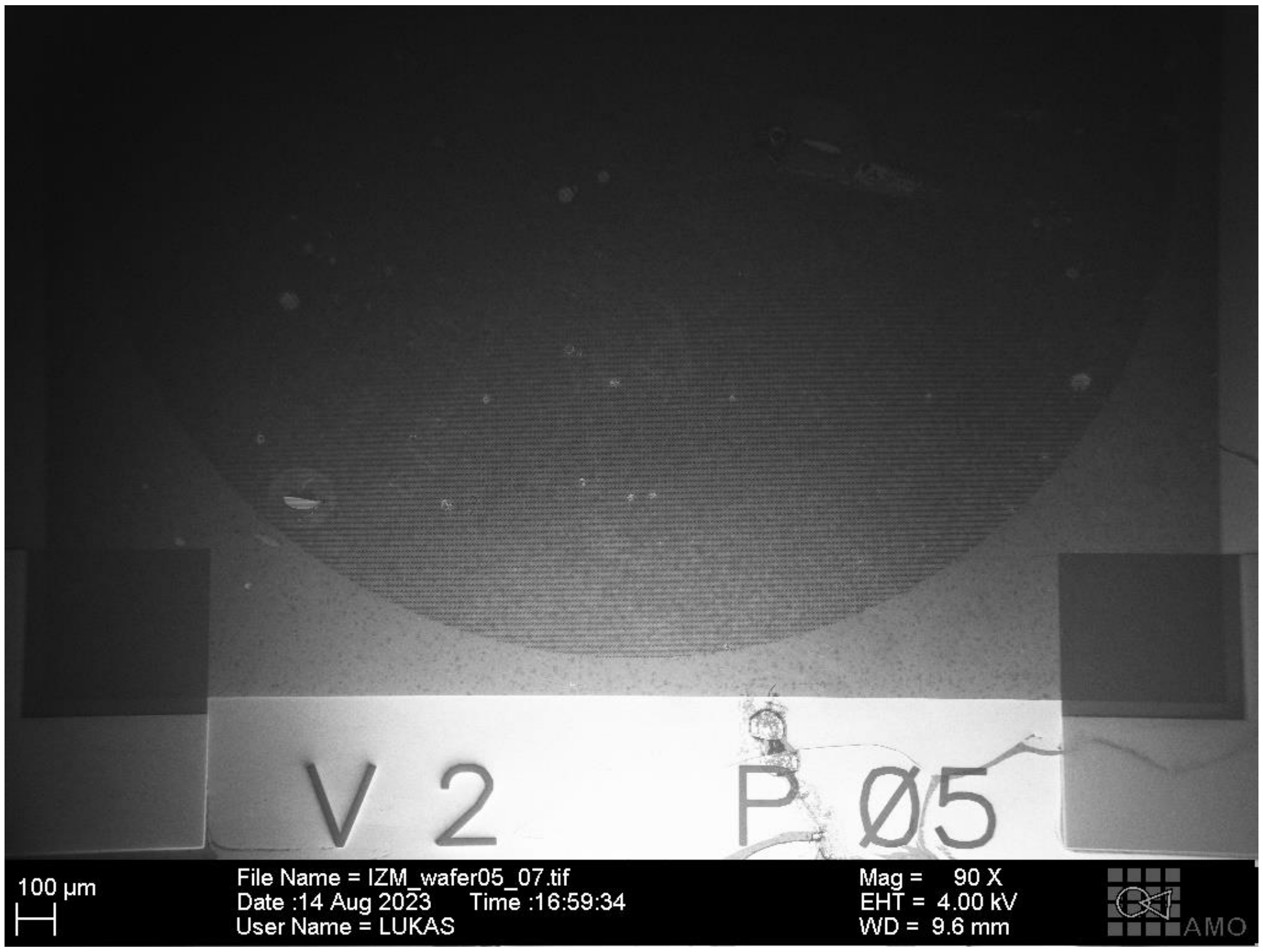

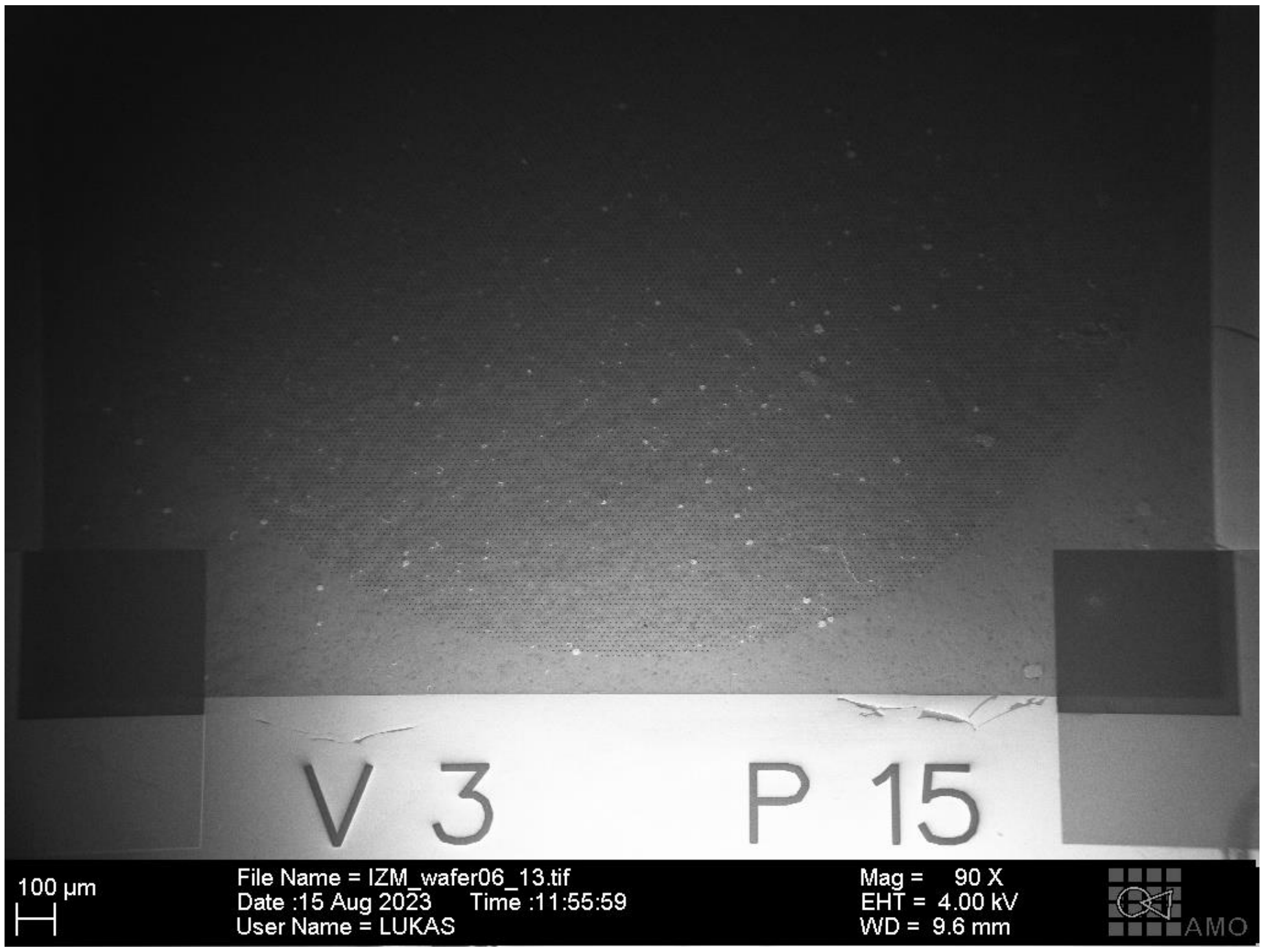

**Figure S3.** SEM images of two large arrays of 2 µm diameter (top) and 3 µm diameter (bottom) double-layer graphene membranes on Fraunhofer IZM substrates. From the whole membrane array of several millimeter width and height, a rectangular area of approximately 1,110 µm × 660 µm per device was imaged at high resolution and evaluated for the membrane yield.

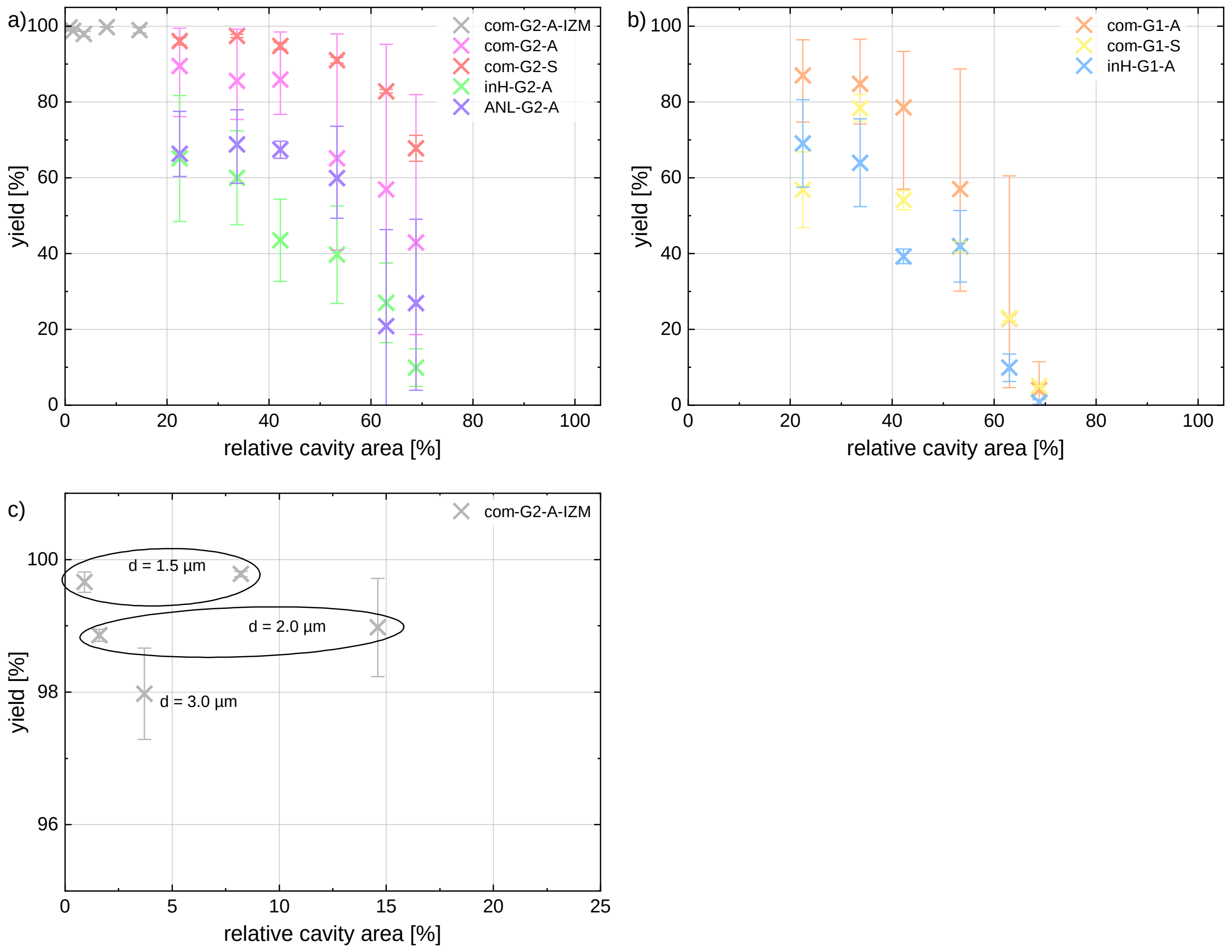

**Figure S4.** Plots of the intact membrane yield against the relative cavity area (*i.e.*, the suspended graphene area in case of 100 % yield) for (a) double-layer and (b) single-layer graphene. (c) Zoom-in into (a) for the Fraunhofer IZM samples.

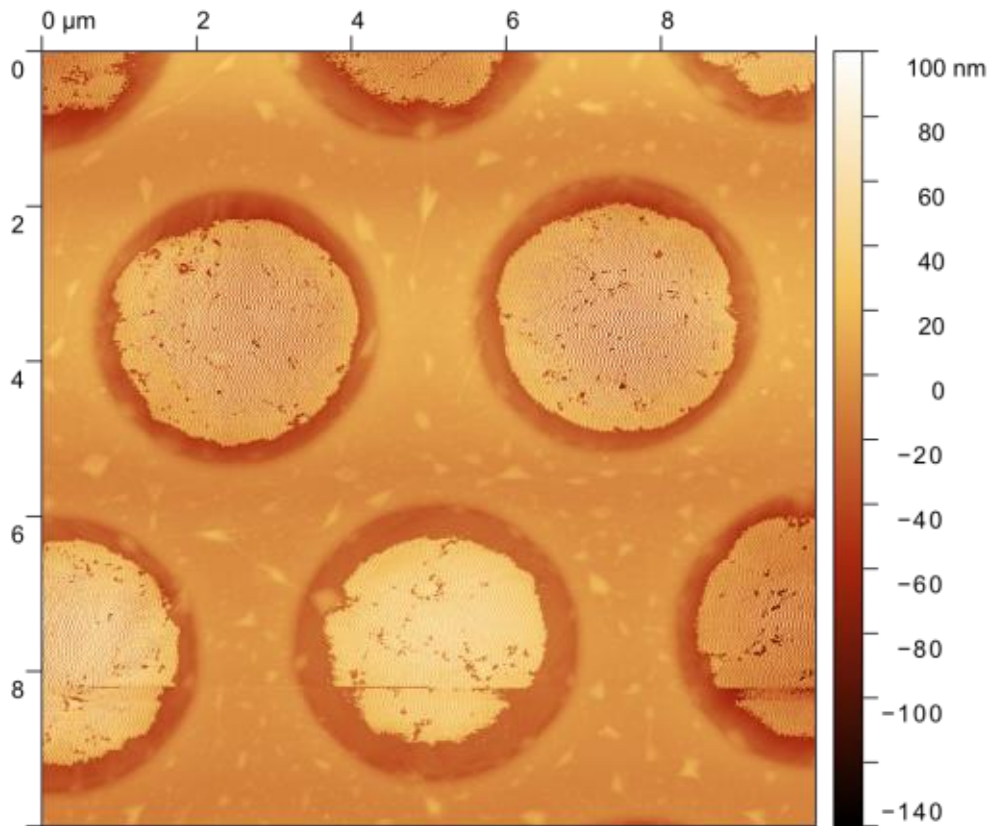

**Figure S5.** AFM scan of sample ANL-G2-A-2.

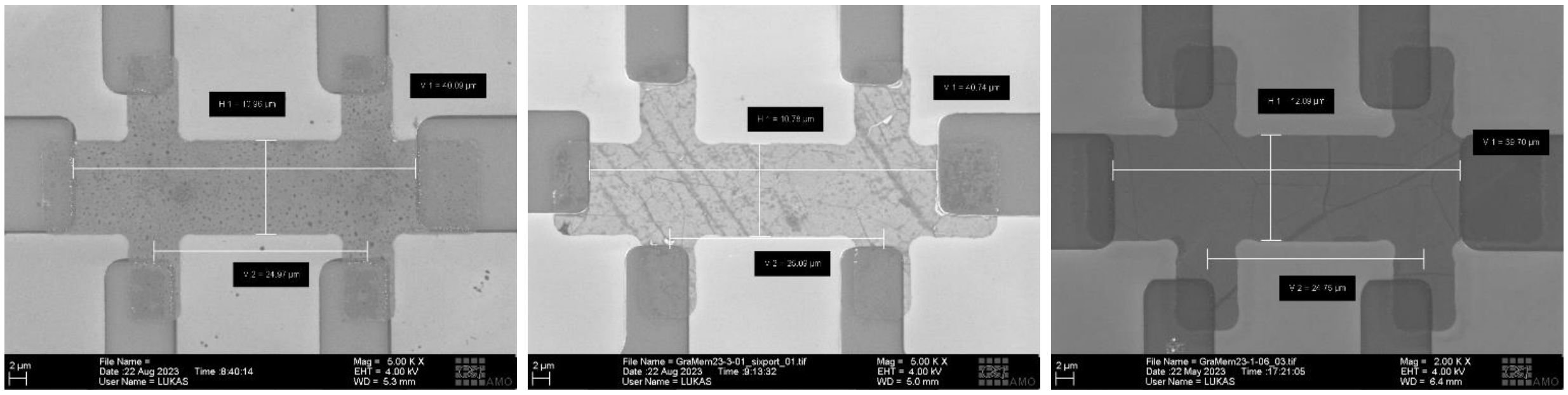

**Figure S6.** SEM images of six-port graphene devices on three different samples (from left to right: ANL-G2-A-3, inH-G1-A-2331, com-G1-A-2316), used for back-gated field-effect measurements and reference measurements inside the pressure chamber set-up.

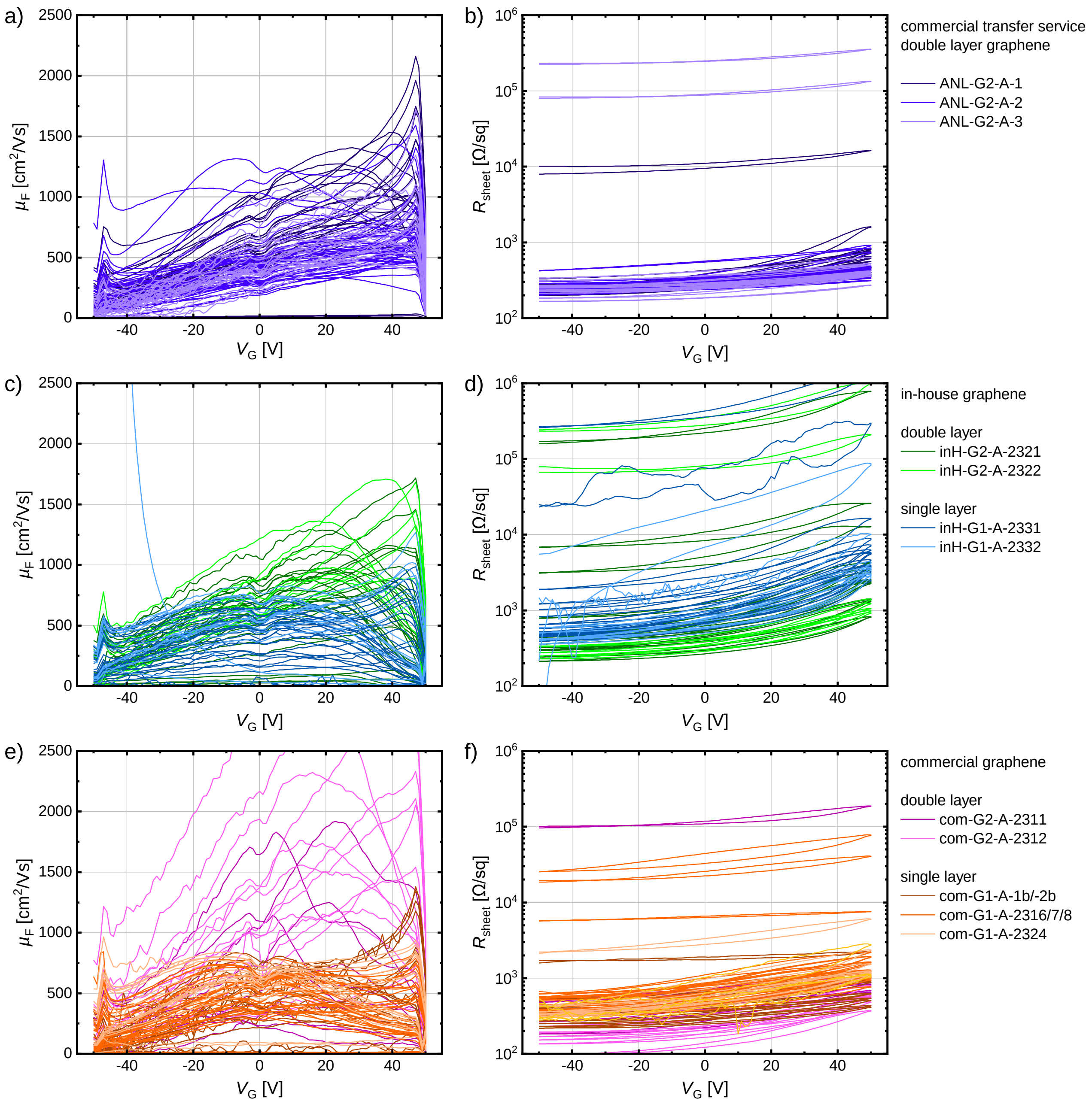

**Figure S7.** Extracted four-probe field-effect mobility and sheet resistance curves.

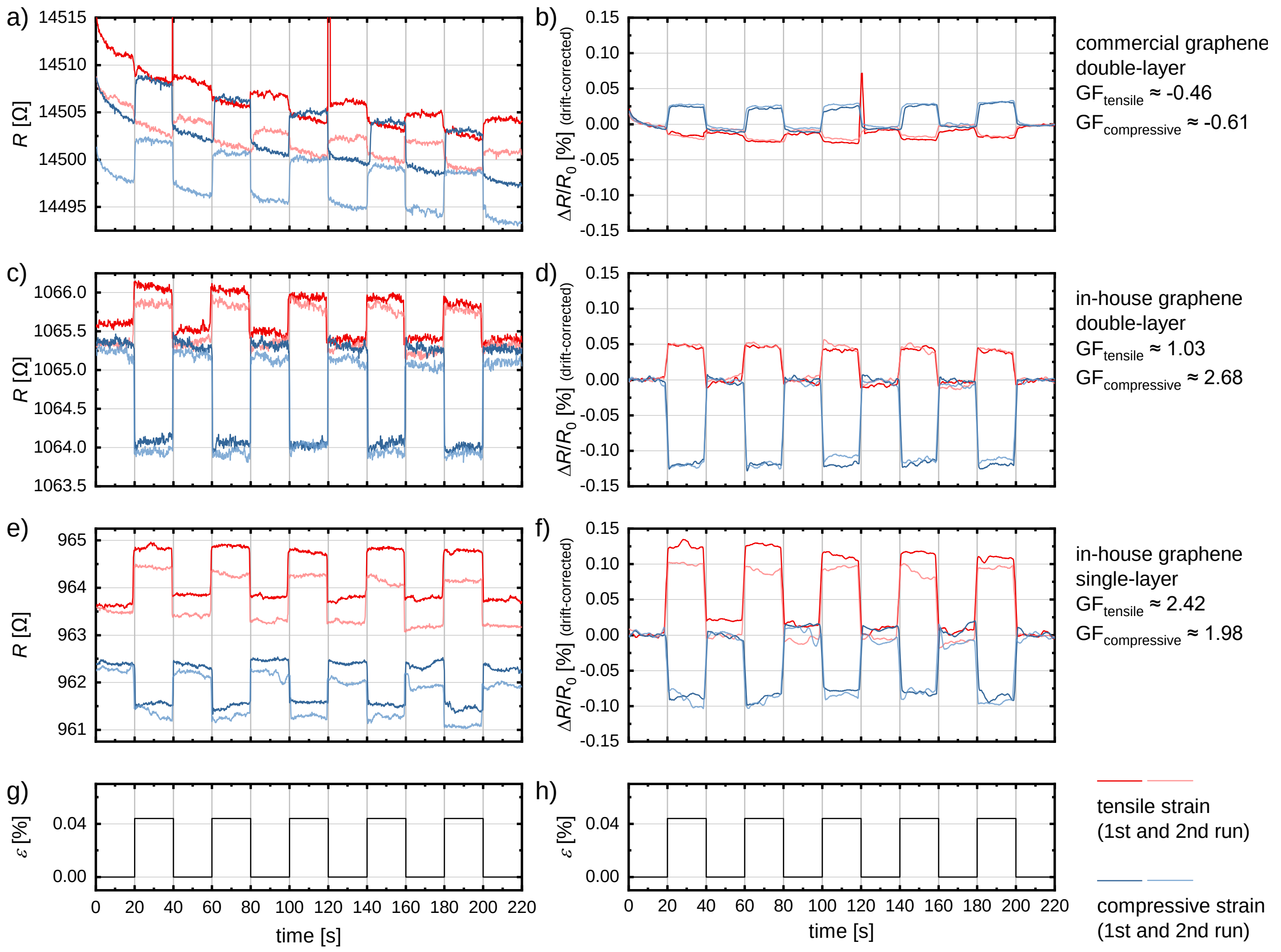

**Figure S8.** Piezoresistive gauge factor (GF) measurements of (a-b) double-layer commercial graphene, (c-d) in-house-grown double-layer graphene, and (e-f) in-house-grown single layer graphene. The samples were glued to either the top surface or the bottom surface of a steel beam, applying either tensile (red color) or compressive (blue color) strain when bending. The approximate strain magnitude is shown together with the time scale in panels (g-h). The extracted GF are shown next to the graphs.

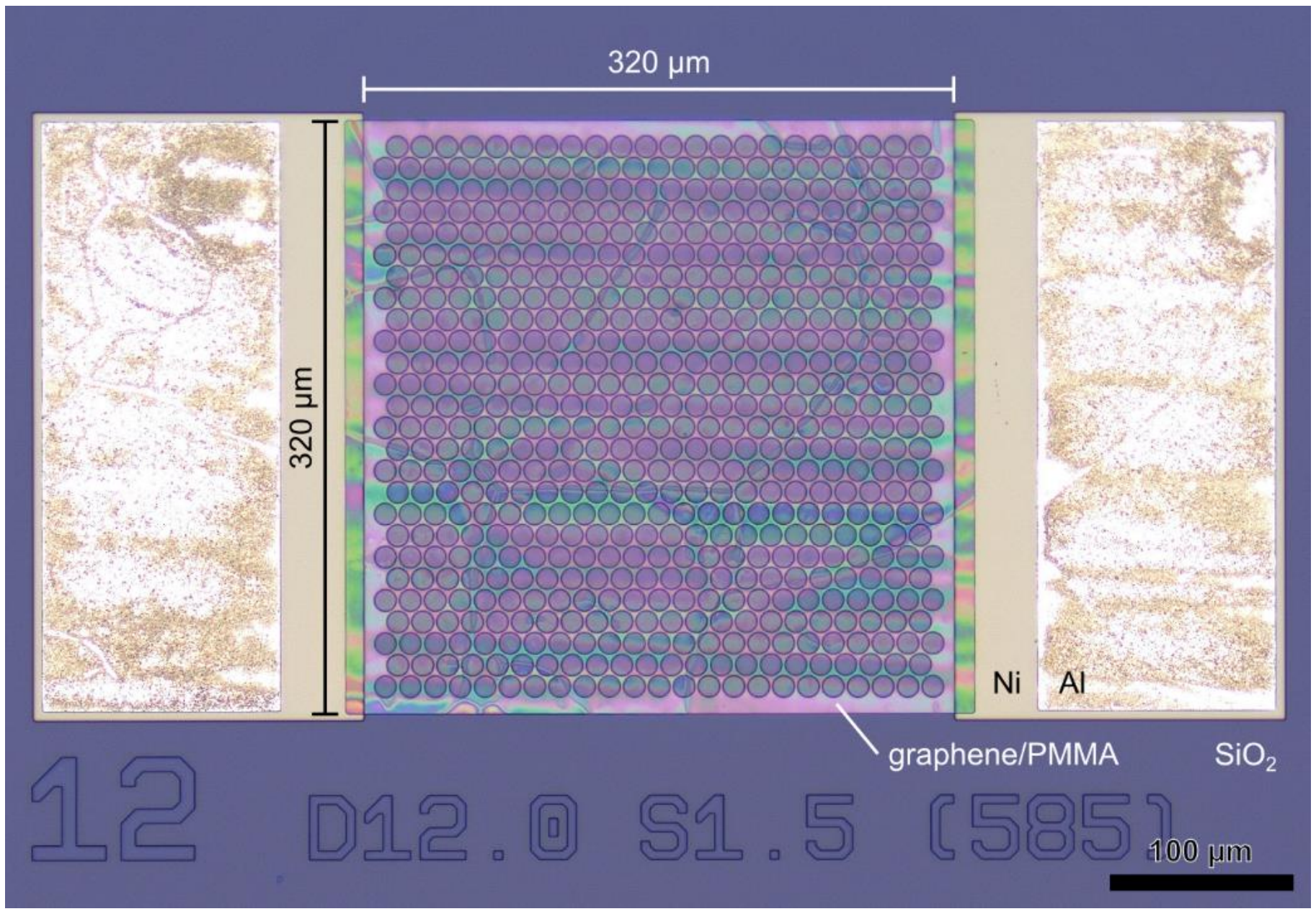

**Figure S9.** Optical micrograph of a pressure sensor device before PMMA removal with device dimensions indicated. The total channel area is 320 µm × 320 µm, while the hexagonal array of membranes is fitted within a 310 µm × 310 µm area. Here, membranes with a diameter of 12 µm are shown.

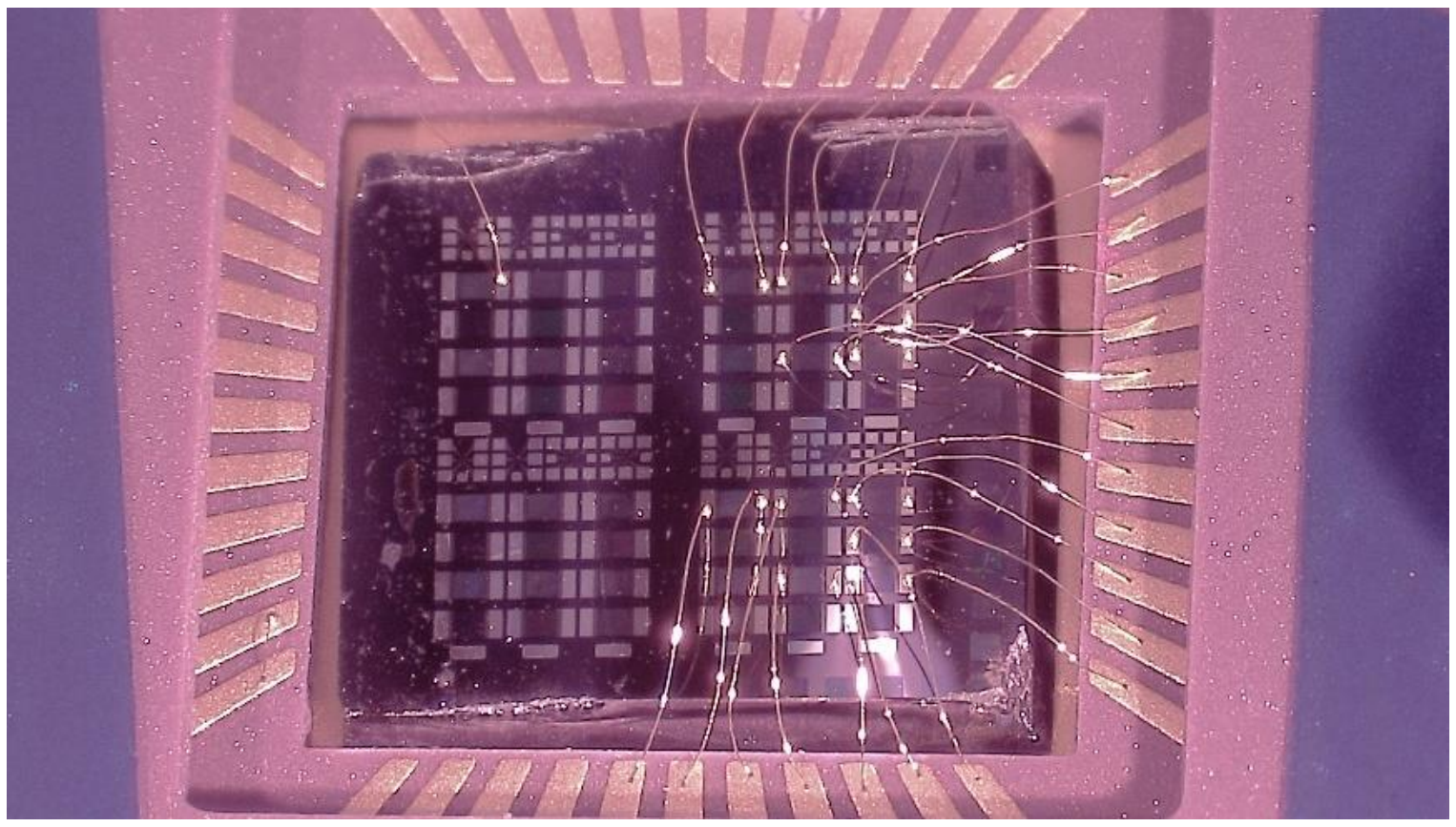

**Figure S10.** Photograph of sample com-G2-A-2d after wire-bonding into chip carrier, before measurements inside the pressure chamber set-up.

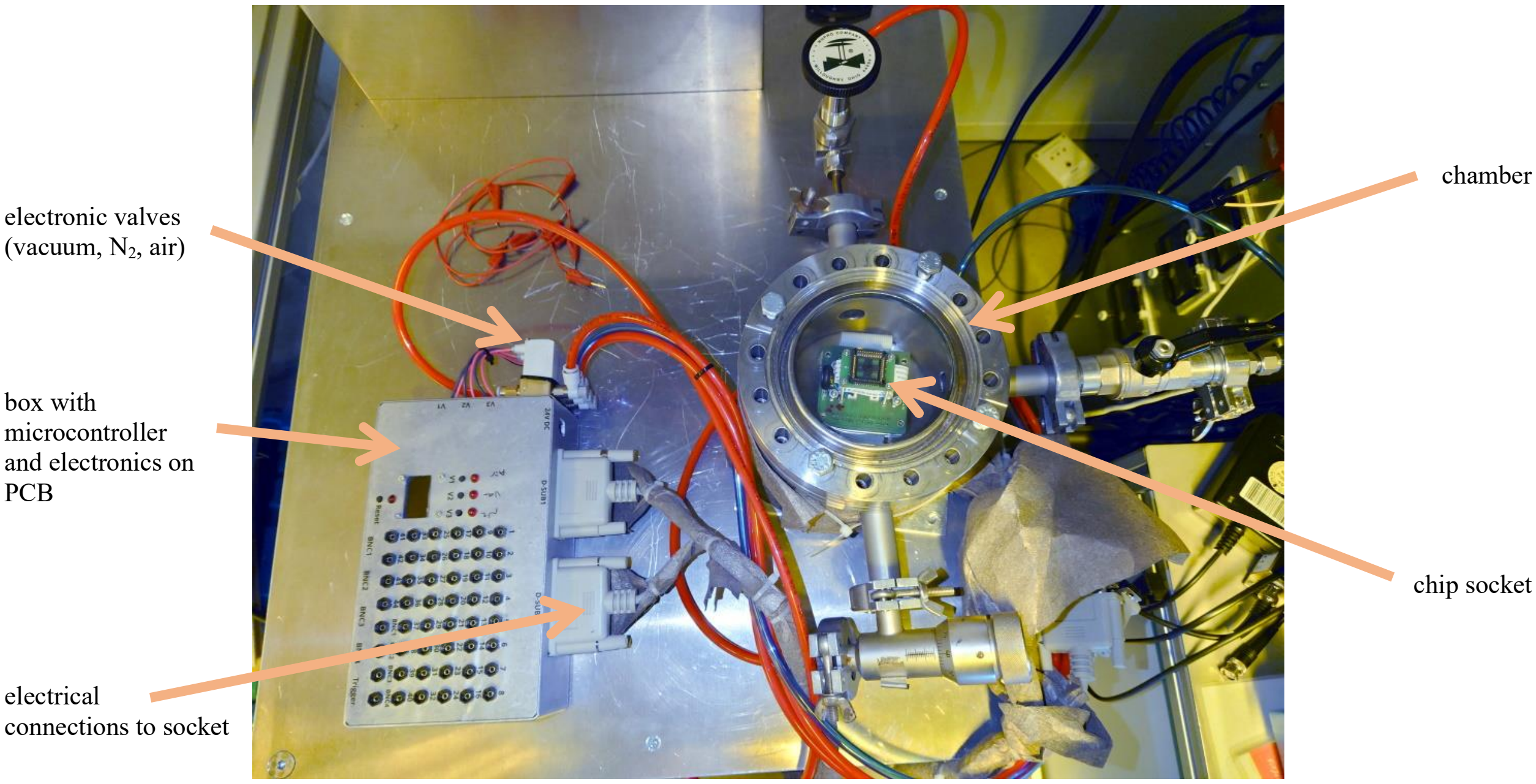

**Figure S11.** Photograph of the pressure chamber set-up used for the electrical measurements of the pressure sensors.